\documentclass[twocolumn]{aastex62}
\pdfoutput=1
\usepackage{hyperref}

\graphicspath{{./}{figures/}}

\shorttitle{Simulating Gas Inflow}
\shortauthors{Melso et al.}
\usepackage{CJKutf8}
\usepackage{threeparttable}
\begin{document}

\title{Simulating Gas Inflow at the Disk-Halo Interface}

\correspondingauthor{Nicole Melso}
\email{nmelso@astro.columbia.edu}

\author{Nicole Melso}
\affil{Department of Astronomy, Columbia University, 550 W. 120th Street, New York, NY 10027, USA}

\author{Greg L. Bryan}
\affiliation{Department of Astronomy, Columbia University, 550 W. 120th Street, New York, NY 10027, USA}
\affiliation{Center for Computational Astrophysics, Flatiron Institute, 162 5th Ave, New York, NY 10010, USA}

\author{Miao Li \begin{CJK*}{UTF8}{gbsn} (李邈) \end{CJK*}}
\affiliation{Center for Computational Astrophysics, Flatiron Institute, 162 5th Ave, New York, NY 10010, USA}



\begin{abstract}
The interaction between inflowing gas clouds and galactic outflows at the interface where the galactic disk transitions into the circumgalactic medium is an important process in galaxy fueling, yet remains poorly understood. Using a series of tall-box hydrodynamic {\sc Enzo} simulations, we have studied the interaction between smooth gas inflow and supernovae-driven outflow at the disk-halo interface with pc-scale resolution. A realistic wind of outflowing material is generated by supernovae explosions in the disk, while inflowing gas is injected at the top boundary of the simulation box with an injection velocity ranging from $10-100 \rm \ km \ s^{-1}$. We find that cooling and hydrodynamic instabilities drive the injected gas to fragment into cold ($\sim 10^{3}$ K) cloud clumps with typical densities of $\sim 1 \rm \ cm^{-3}$. These clumps initially accelerate before interacting and partially mixing with the outflow and decelerating to velocities in the 50-100 $\rm km \ s^{-1}$ range. When the gas clumps hit the disk, $10\%-50 \%$ of the injected material is able to accrete (depending on the injection velocity). Clumps originating from gas injected with a higher initial velocity approach the disk with greater ram pressure, allowing them to penetrate through the disk in low density regions. We use (equilibrium) {\sc Cloudy} photoionization models to generate absorption and emission signatures of gas accretion, finding that our mock HI and H$\alpha$ observables are prominent and generally consistent with measurements in the Milky Way. We do not predict enhanced emission/absorption for higher ionization states such as OVI. 

\end{abstract}


\section{Introduction} \label{sec:intro}
Gas accretion is needed throughout the cosmic evolution of spiral galaxies in order to replenish the galactic disk with hydrogen and fuel long-term star formation \citep{Erb2008,Hopkins2008,Putman2009}. Without a fresh influx of low metallicity gas \citep{Chiappini2001,Fenner2003,Sommer2003}, most star-forming galaxies including the Milky Way will exhaust their supply of disk gas \citep{Chomiuk2011,Leroy2013}, although stellar outflows can delay this \citep{Leitner2011}. 

There are a variety of external gas reservoirs beyond the galactic disk that may contain ample gas for sustaining a constant star formation rate. The majority of low-redshift baryons lie in the intergalactic filaments that make up the large-scale structure \citep{1999ApJ...514....1C, 2012ApJ...759...23S}. These filaments feed into the potential wells around galaxies, where circumgalactic gas pools to form the galactic halo. Due to its proximity to the galactic disk, this circumgalactic medium (CGM) is a direct source of fuel for future star formation and closely influences galaxy evolution \citep{Putman2012, Tumlinson2017}. The CGM is a multi-phase combination of inflowing material from the intergalactic medium (IGM), gas stripped from orbiting satellite clouds, and enriched outflow expelled from the galactic disk. All of these sources are thought to fuel the galactic disk but with differing accretion processes \citep{Putman2017}.

Cosmological simulations have demonstrated that gas enters into the CGM along the cosmic web, resulting in filamentary inflow within the hotter, more diffuse halo gas. This inflow can be primarily cold when the halo virial temperature is relatively low (and so cooling is effective) and/or at high redshift, when the filament density is higher \citep[e.g.,][]{Keres2005}. The inflow transitions to hotter gas flows at higher masses and lower redshifts. There is evidence from recent simulations that such flows heat, either via shocks or mixing before entering the galaxy \citep{Nelson2013}, although the numerical resolution in such cosmological simulations tends to be relatively poor in the CGM. Cosmological numerical models of halos with masses similar to the Milky Way suggest that low-redshift accretion is dominated by warm-hot ($10^{5} < $ T $< 10^{6}$ K) filamentary inflow at $50-150$ km s$^{-1}$ \citep{Joung2012}; such inflow fragments and mixes with feedback from the galactic disk \citep{Joung2012, 
Fernandez2012}. 

The majority of inflowing gas is expected to accrete towards the outskirts of a galaxy, avoiding strong feedback form the galactic center \citep{Stewart2011,Fernandez2012}. Even towards the outskirts of a galaxy, gas will encounter stellar feedback on its path to the disk. In cosmological hydrodynamic simulations, interactions between filamentary inflow and the surrounding halo/outflowing gas are important for creating over-densities in the filaments that facilitate gas cooling and fragmentation \citep{Keres2009,Joung2012,Fernandez2012}. Interactions between inflowing and outflowing gas at the disk-halo interface are likely similarly important in the final stages of gas accretion, but are difficult to study with the kpc-scale resolution in cosmological zoom-in hydrodynamic simulations.

All gas exchanged between the galaxy and the surrounding halo must pass through a boundary layer where the galactic disk transitions into the lower galactic halo. This region is called the disk-halo interface \citep{Norman1989, Ford2010, Putman2012}, and it is a crucial boundary for studying the final stages of gas accretion. Across the disk-halo interface, there is a transition in the phase, density, and kinematic properties of the gas. A layer of dense, cold HI gas makes up most of the disk mass in the solar neighborhood, ranging in thickness from $<100$ pc in the galactic  center to  $\sim 1$ kpc along the Galactic Warp \citep{Dickey1990,Levine2006}. This neutral component transitions into a diffuse warm ionized component (or Reynolds Layer) extending to a height of $\sim 2$ kpc \citep{Reynolds1993,Haffner2003,Gaensler2008}. An extensive population of cold HI clouds have been found  embedded in this warm gas layer within $1-2$ kpc of the galactic disk \citep{Lockman2002,Ford2010, Peek2011, Saul2012}. The volume above the diffuse warm ionized layer is filled by warm-hot/hot gas blending into the Galactic halo \citep{Miller2013,Miller2015}

\begin{table*}
\caption{Summary of gas injection parameters in each simulation.} % title of Table
\centering % used for centering table
\begin{tabular}{c c c c c c c c} % centered columns (4 columns)
\hline\hline %inserts double horizontal lines
Name & Velocity [km s$^{-1}$] & Density [$\rm cm^{-3}$] & Temperature [K] & Start Time [Myr] & Stop Time [Myr] & Mass$^{\dagger}$ [g] \\ [0.5ex] 
%heading
\hline % inserts single horizontal line
BURST50 & 50 & 0.05 & $10^{4}$ & 50 & 75 & $\rm 2.10 \times 10^{5} M_{\odot}$ \\ % inserting body of the table
BURST100 & 100 & 0.05 & $10^{4}$ & 50 & 62 & $\rm 2.11 \times 10^{5} M_{\odot}$ \\
IMD10 & 10 & 0.05 & $10^{4}$ & 0 & 70 &$\rm 1.04 \times 10^{5} M_{\odot}$   \\ [1ex] % [1ex] adds vertical space
\hline %inserts single line
\end{tabular}
\begin{tablenotes}\footnotesize
\item $^{\dagger}$ Although the density and size the injected cloud is constant between runs, the total mass of the inflowing gas cloud is dependent on the deceleration experianced upon impact with the outflowing material.
\end{tablenotes}
\label{table:parameters}
\end{table*}

The gas distribution at the disk-halo interface is likely to be quite dynamic, due to multiple effects. This includes the impact of feedback expelling hot gas to large scale heights as well as possible gas cooling close to the galactic disk \citep{Joung2006, Hill2012, Putman2012,Creasey2013,Li2017}. Additional evidence of gas cooling is provided by the observed HI cloud populations associated with the disk-halo interface \citep{Lockman2002, Ford2010, Peek2011, Saul2012}. These clouds are thought to condense out of the surrounding medium, and a variety of models have been proposed to facilitate this gas cooling \citep{Bregman1980, Norman1989, Houck1990, Fraternali2008, Marasco2012, Joung2012b}. The survival of these clouds, and therefore their ability to fuel the galactic disk, is dependent on their mass, their height above the disk, and their interactions with the surrounding gas as well as galactic outflow \citep{Heitsch2009}.

Extragalactic observations around spiral galaxies reveal multiple pieces of evidence supporting the presence of gas inflow. One of the strongest pieces of direct evidence for accretion comes from absorption line spectra of UV-bright stars across the disk of M33. Since M33 is inclined at $55^{\circ}$, these observations are able directly probe the vertical gas kinematics, revealing disk-wide ionized gas inflow \citep{Zheng2017}. At greater distances, the disk-halo interface has been observed in handful of edge-on extragalactic systems. These observations probe the vertical velocity structure of the gas indirectly through measurements of the radial velocity profile. Gas in the halos of edge-on galaxies exhibits a kinematic transition where both neutral gas \citep{Sancisi2001, Fraternali2002, Oosterloo2007,Heald2011} and ionized gas \citep{Rand2000, Heald2006, Heald2007, Bizyaev2017} decrease in rotational velocity as a function of distance from the galactic disk. This rotational velocity gradient suggests the presence of gas inflow at the disk-halo interface, as inflowing gas should lag the rotational velocity of the disk and accreting gas merging with the galactic disk should be co-rotating. In the Milky Way, this trend is observed in the the populations of discrete intermediate velocity HI clouds observed at the disk-halo interface, which are known to exhibit net inflow velocities \citep{Wakker2001, Lockman2002, Ford2010, Peek2011, Saul2012}. In a recent sample, these HI clouds can be split into a kinematically warm population lagging the rotation of the disk at larger scale heights and a kinematically cold co-rotating population within 1 kpc \citep{Saul2012}. 

The presence of outflowing gas at the disk-halo interface regulates gas accretion and possibly even re-fuels the galactic disk in a fountain cycle of expulsion and re-accretion \citep{Shapiro1976, Bregman1980}. Multiple feedback mechanisms expel gas from the galactic disk, influencing the evolution of the galaxy as well as the composition of the circumgalactic medium. Radiative feedback sends ionizing photons into the halo while mechanical feedback adds material, energy, and momentum from the disk back into the surrounding CGM. One of the most energetic and ubiquitous forms of mechanical feedback is stellar feedback in the form of overlapping supernovae explosions \citep{Joung2006,Creasey2013,Fielding2018}, which can launch multi-phase outflows of metal-rich gas into the CGM with velocities of $\sim 1 - 100$ km s$^{-1}$ \citep{Li2017,Kim2017}.

We present a suite of small box hydrodynamic {\sc ENZO} simulations designed to study the interaction between inflowing and outflowing gas at the disk-halo interface with pc-scale resolution. This work builds upon simulations of gas outflow presented in \cite{Li2017}, where they studied the energy, mass, and metallicity of gas blown out of the galactic disk by overlapping supernovae explosions. We have added gas inflow to these simulations, the setup for which is described in Section 2. Section 3 presents the results of our simulation suite, and an overview of the gas evolution, cloud fragmentation, and eventual accretion outcomes. We use this simulation suite in conjunction with photoionization modeling to make observational predictions in Section 4. Finally, we discuss the implications of this work and and summarize our findings in Sections 5 and 6 respectively.

% -----------------------------------------

\section{Methods}\label{sec:methods} 
The simulations in this paper were performed using the adaptive-mesh hydrodynamics code {\sc Enzo} \citep{2014ApJS..211...19B}. Each simulation is a small box centered on the galactic plane with supernovae-driven outflow throughout the simulation volume. The initial setup as well as the supernova outflow prescription are identical to the $\rm \Sigma 10-KS$ model detailed in \cite{Li2017}. We refer to that paper for details, but also briefly describe the initial setup in Section~\ref{sub:setup} below. In this work we have augmented that work by injecting inflowing gas at the top boundary of the box in order to study the interaction between inflowing and outflowing gas at the disk-halo interface; inflow characteristics are described in more detail in Section~\ref{sub:inflow}.

\subsection{Initial Setup} \label{sub:setup}
These simulations focus on a local volume of gas centered on the galactic disk. Each simulation is initialized as a rectangular box (elongated along the z-axis) with [x,y,z] dimensions $[0.35 \times 0.35 \times 5]$ kpc. A disk of gas is set up in the x-y plane at the center of the box (z = 0), and a stratified density halo extends $\pm 2.5$ kpc above and below the disk (with an initial temperature of $10^4$ K). The x and y boundaries of the box remain periodic throughout the simulation, and the z-axis boundaries are initially set to outflow. After a time delay (given in Table \ref{table:parameters} under ``Start Time''), the top z-axis boundary is changed to inflow and we inject a 1 kpc-long filament of gas inflowing towards the galactic disk. We use adaptive mesh refinement (AMR) to resolve the detailed interactions between the inflowing and outflowing gas. There are two levels of grid refinement within the simulation volume, one at $\pm 1$ kpc from the disk and another at $\pm 0.5$ kpc. Each sub-grid increases the resolution by a factor of two. The finest grid within $\pm 0.5$ kpc of the galactic disk achieves a spatial resolution of 2 pc.

We model the galactic disk near the solar neighborhood with a gas surface density ($\rm \Sigma_{gas}$) of $\rm 10 \ M_{\odot} \ pc^{-2}$, corresponding to a SFR surface density ($\rm \dot{\Sigma}_{SFR}$) of $6.31 \times 10^{-3} \rm \ {M}_{\odot}^{-1}$ \ kpc$^{-2}$ yr$^{-1}$ on the Kennicutt-Schmidt (KS) relation \citep{1989ApJ...344..685K}. The photoelectric heating rate is set to a constant value of $\rm 1.4 \times 10^{-26} \ ergs \ s^{-1}$ per H atom. The gas is initially isothermal and distributed in hydrostatic equilibrium, with gravity modeled as the combination of a baryonic stellar disk and an NFW dark matter halo. We apply a gas cooling curve appropriate for half solar-metallicity gas, ranging in temperature from $300 - 10^{9}$ K \citep{Rosen1995, Tasker2006}.

In order to drive inhomogeneous outflows, supernovae are injected near the midplane. Each supernova adds $10 \ M_{\odot}$ of gas and $10^{51}$ ergs of thermal energy. The horizontal (x-y) distribution of supernovae is random. We implement both core-collapse and Type Ia supernovae using two different vertical distributions. Most of the supernovae ($90 \%)$ are core-collapse, modeled as a gaussian vertical (z) distribution with a scale height $\rm h_{cc} = 150$ pc. The remaining $10 \%$ are Type Ia supernovae distributed exponentially along the z-axis with a scale height $\rm h_{Ia} = 325$ pc. Supernovae are set off at a constant rate of $\dot{\Sigma}_{SFR}/m_{0} = \rm 4.2 \times 10^{-5} \ kpc^{-2} \ yr^{-1}$, where we assume one SN explosion for every $m_{0} = 150 \ M_{\odot}$ of star formation. This supernovae prescription drives material towards the outflow z-axis boundaries of the simulation box. See \cite{Li2017} for more details of the initial setup as well as a detailed analysis of the resulting outflows.

\begin{figure}[t]
\centering
\includegraphics[width=\columnwidth]{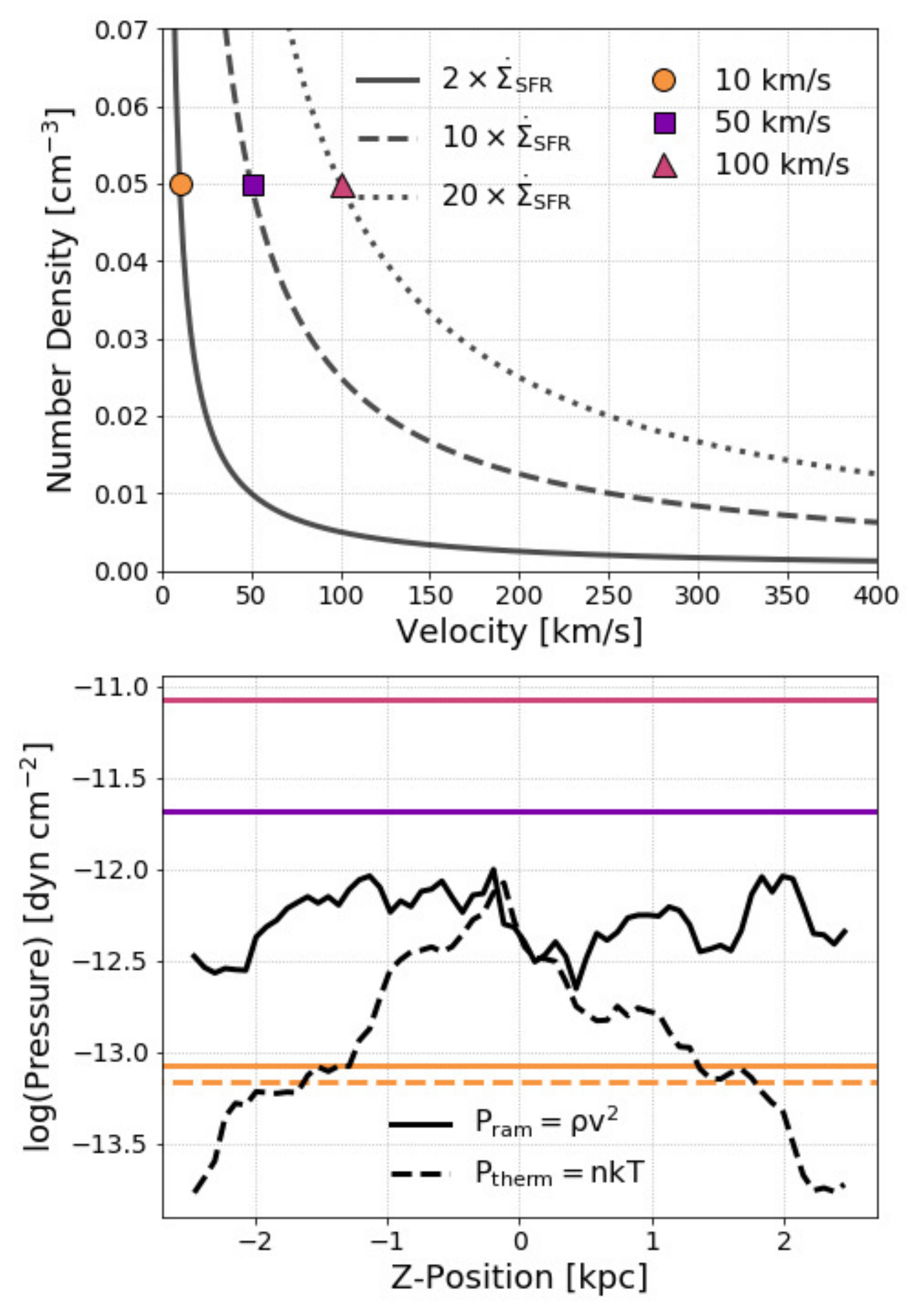}
\caption{(Top) Isopleths of the inflowing gas mass flux in equilibrium with the star formation rate surface density $\rm \dot{\Sigma}_{SFR}$ (solid), 10$\rm \times \dot{\Sigma}_{SFR}$ (dashed), and 20$\rm \times \dot{\Sigma}_{SFR}$ (dotted). The orange circle, purple square, and pink triangle mark the initial density and inflow velocity conditions used in our three simulation runs. (Bottom) 1D profiles of the volume-weighted ram pressure (black solid) and thermal pressure (black dashed) of the steady-state supernovae-driven outflow as a function of z-position. The ram pressure of inflowing gas in three simulation runs is shown in solid colors corresponding to the labels in the top panel. All three inflow simulations have the same thermal pressure shown as a dashed orange line. }
\label{fig:initial}
\end{figure}

\subsection{Inflow Conditions} \label{sub:inflow}
In order to inject a cloud into the domain, the top z-axis boundary condition is changed from outflow to inflow. This is generally done after a time delay (see Table \ref{table:parameters}) in order to allow the outflow to become established. Supernovae continue to expel gas towards the boundaries, but this outflow is now met by a layer of inflowing material. All injected gas at the boundary is given an initial density, temperature, and z-velocity towards the disk. The initial values for these injected gas parameters are set by a number of considerations, including the desired inflow mass flux as well as the ram pressure of the outflowing gas. We characterize the inflow rate in terms of the star formation rate surface density, motivated by the idea that the inflowing gas should more than replenish the ongoing star formation in the disk. Accordingly, we compare the injected mass flux ($\rho v$) with the star formation rate surface density ($\rm \dot{\Sigma}_{SFR}$). For example, if inflows of a given rate had a (temporally and spatially averaged) covering fraction of $10\%$, then the inflow rate (when ``on") required to maintain star formation, on average, would be $10$ times the mean star formation rate surface density. Based on this logic, we explore ranges from $2-20$ times the mean star formation rate surface density. An additional constraint is that the ram pressure of the injected gas must be greater than the ram pressure of the outflowing gas. If this condition is not met, the injected gas will be unable to enter the box, an issue which we discuss more in the following explanation of Figure \ref{fig:initial}.

Figure~\ref{fig:initial} (top) shows the combinations of gas density and z-velocity resulting in an injected mass flux equivalent to $2 \times \rm \dot{\Sigma}_{SFR}$ (solid), $10 \times \rm \dot{\Sigma}_{SFR}$ (dashed), and $20 \times \rm \dot{\Sigma}_{SFR}$ (dotted). The orange circle, purple square, and pink triangle denote the initial values used in each of our three simulation runs (See Table\ref{table:parameters}); they fall on the three given mass flux isopleths respectively. Figure \ref{fig:initial} (bottom) shows 1D profiles of the ram pressure (black solid) and thermal pressure (black dashed) of the outflowing gas at 50 Myr. By this time, the supernovae explosions have set up a continuous (steady-state) stream of inhomogeneous outflow at the z-axis boundaries. At these boundaries, the ram pressure ($\rm P_{ram} = \rho v^{2}$) of the steady-state outflowing gas dominates the total pressure ($\rm P_{ram} + P_{thermal}$). The ram pressure of the injected gas for each simulation is shown in solid colors corresponding to the legend in Figure \ref{fig:initial}a. The ram pressure of the 50 km s$^{-1}$ and 100 km s$^{-1}$ injected gas exceeds the ram pressure of the outflowing gas, allowing injected gas with these initial conditions to enter the box at 50 Myr. Gas injected at 10 km s$^{-1}$ comes close to equating the mass flux with the star formation rate surface density ($\rho v = \rm \dot{\Sigma}_{SFR}$),  but the ram pressure of this low-velocity injected gas falls a factor of 3 below the ram pressure of the outflowing gas. As described in Section \ref{sub:sim}, we manage to simulate this low-velocity inflow by immediately injecting gas into the simulation box at the first time step, when the outflow ram pressure is much lower. 

In addition to setting the initial density, temperature, and z-velocity of the injected gas, we have also added a colour density field to tag the injected gas and track its progression throughout the simulation volume. The colour density is the density of only the injected gas (from the top boundary), while the rest of the gas in the domain starts with a near-zero colour density field (including SN injected gas in the mid-plane). We can divide the colour density by the total gas density to create a normalized colour density field, which represents the fraction of inflowing gas present in a cell volume. In our analysis we use both the colour density and the normalized colour density. In some analysis, we use a normalized colour density cut of $>50\%$ to separate the properties of the injected gas from those of the outflowing gas. 

\subsection{Simulation Overview} \label{sub:sim}

We have run three variations of our basic simulation: BURST50, BURST100, and IMD10. Each follows the initial setup described in Section~\ref{sub:setup} but with different inflow rates. The number in the naming scheme for each simulation is the z-velocity of the injected gas in km s$^{-1}$. The injected gas is given an initial density (n = 0.5 $\rm cm^{-3}$) and temperature (T = $10^{4}$ K), both of which are kept constant across all runs. The runs are differentiated by the start time of the gas injection, the z-velocity of the injected gas, and the injection stop time. The injection duration (stop $-$ start) is set so that the inflowing gas in each simulation reaches a length of $\sim 1$ kpc, corresponding approximately to a mass of $10^5$ M$_\odot$. We halt the gas injection by reverting the top z-boundary condition to outflow, creating a finite burst of injected gas inflowing towards the galactic disk. 
 
The BURST50 and BURST100 simulations inject a burst of inflowing gas at 50 km s$^{-1}$ and 100 km s$^{-1}$ respectively. Both of these runs simulate inhomogeneous supernovae-driven outflows interacting with an inflowing $\sim 1$ kpc-long burst of injected gas. We delay the injection start time by 50 Myr in order to set off enough supernovae to develop a steady-state outflow at the box boundaries. The IMD10 run immediately injects gas at 10 km s$^{-1}$ starting from the first time step. We are able to lower the inflow velocity in this run because steady state outflow has not yet been established at early times, the outflow ram pressure is much lower, and the 10 km s$^{-1}$ injected gas enters the box relatively unopposed. The injected gas is initially only in contact with the halo and later impacts the outflowing gas closer to the disk. We again stop injecting gas when the 10 km s$^{-1}$ inflow reaches about 1 kpc in length. 

% -----------------------------------------

\section{Results} \label{sec:results}
We present the results of our three primary simulations: the BURST50, BURST100, and IMD10 runs. This analysis was performed using the python package yt \citep{Turk2011}. We begin with an overview of the physical gas structure (Section~\ref{sub:physical}) and the time evolution throughout each simulation (Section~\ref{sub:global}). We then detail the fragmentation of the injected gas (Section~\ref{sub:clouds}) and its ultimate fate upon reaching the disk (Section~\ref{sub:fate}). 

\begin{figure*}%[t]
\centering
\includegraphics[width=\textwidth]{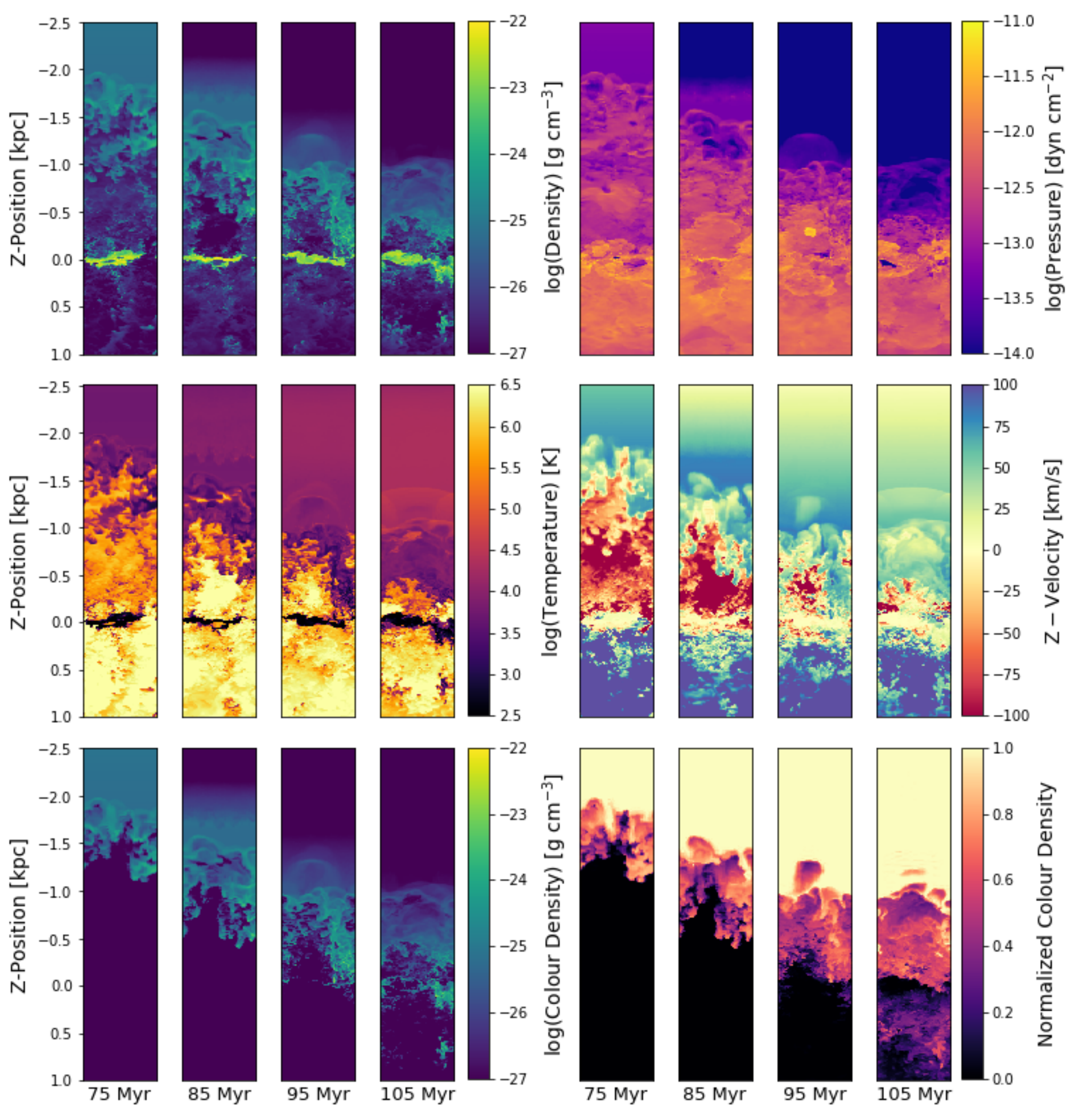}
\caption{Density, pressure, temperature, z-velocity, colour density, and normalized colour density slices showing the time evolution of the BURST50 simulation. Each panel is a slice in the y-z plane at x = 0 with dimensions [0.35 x 3.5] kpc. Note that z-axis is vertical and each slice is only showing the top $70\%$ of the box, as the full simulation is 5 kpc in the z-direction. The time step for each panel is noted along the bottom of the plot. The plane of the disk is the yellow high-density region at z = 0. Gas is injected at the top of the box, and can be seen between z $\approx -2.5$ kpc to z $ \approx -1.5$ kpc at the first time step. This injected gas is given an initial density of $8.35 \times 10^{-26} \rm \ g \ cm^{-3}$, an initial temperature of $10^{4}$ K, and an initial velocity of 50 km $\rm s^{-1}$. The colour density traces the density of only the injected gas, while the rest of the gas in the domain starts with a near-zero colour density field. We divide the colour density by the total gas density to create a normalized colour density field, which represents the fraction of injected gas present in a cell volume. We have changed the aspect ratio in these slices (compressing them along the z-direction) to ensure they fit in this figure. Movies (with the correct aspect ratio) for all three simulations showing the frame-by-frame time evolution can be found in the appendix}. 
\label{fig:panels}
\end{figure*}

\subsection{Physical Structure} \label{sub:physical}

We begin by briefly describing how the inflowing, high density gas reacts as it impacts the outflowing, lower-density material. This situation produces the classic three-wave structure familiar from supernovae studies. When we inject gas at the top of the simulation box, we create a contact discontinuity at the boundary between the outflowing and injected material, as well as a weak ``forward" shock propagating down into the outflowing gas, and a relatively high-density ``reverse" shock of decelerated injected gas, which piles up behind the cloud front. This reverse shock propagates up into the inflowing wind (which is assumed to be relatively cool) but at a speed lower than the inflow rate so that the reverse shock is swept down towards the galaxy.

This structure can initially be clearly seen at the upper boundary condition immediately after injection (not shown), but rapidly evolves into a more complicated mixture of clouds as instabilities set in. As is well known, higher density gas above lower density gas at a contact discontinuity in a gravitational field creates a boundary that is Rayleigh-Taylor (RT) unstable. The reverse shock in the injected gas amplifies the growth of the RT instability by increasing the density disparity at the contact discontinuity. Rayleigh-Taylor fingers form from the cold, high-density injected gas. Over time, these features grow in length, and shearing with the outflowing gas triggers the Kelvin-Helmholtz (KH) instability. The RT and KH instabilities in combination lead to the eventual fragmentation of the injected gas into smaller clumps (Section \ref{sub:clouds}) that will either mix into the outflowing gas, pass through the disk, or successfully accrete (Section \ref{sub:fate}). This rapid cloud formation means that the precise form of injected gas is unlikely to dramatically change the results we find below.

\subsection{Global Evolution} \label{sub:global}

\begin{figure*}
\centering
\includegraphics[width=\textwidth]{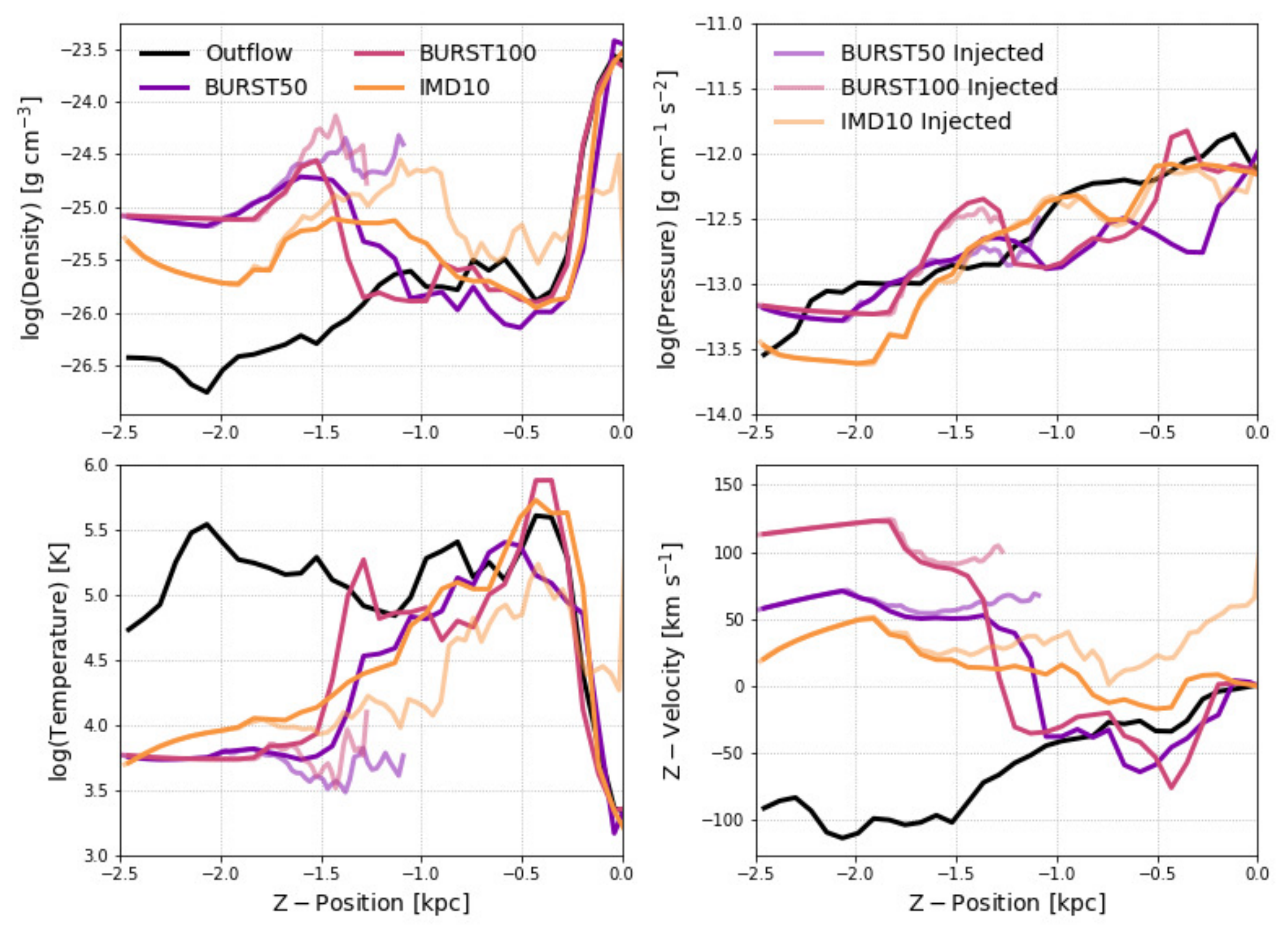}
\caption{1D profiles of the average density, temperature, pressure, and z-velocity as a function of z-position in each of three simulation runs. The density and pressure profiles are weighted by volume, and the temperature and z-velocity profiles are weighted by mass. A negative z-position is above the disk and a positive z-position is below the disk. The opaque (darker) profile lines are averaged over all of the gas in the simulation box. The transparent (lighter) profile lines are averaged over only the injected gas. We separate inflowing gas from the total gas content using the normalized colour field, which represents the fraction of injected gas present in a cell volume. We apply a normalized colour cut of $> 50 \%$. The profile of the outflowing gas (black) is taken at the last time step before inflowing gas is added at the top boundary. The profiles of the three simulation runs are taken after the $\sim 1$ kpc long inflow burst has entered the box at the last time step before the top boundary is switched back to outflow. }
\label{fig:profiles}
\end{figure*}

\begin{figure*}
\centering
\includegraphics[width=\textwidth]{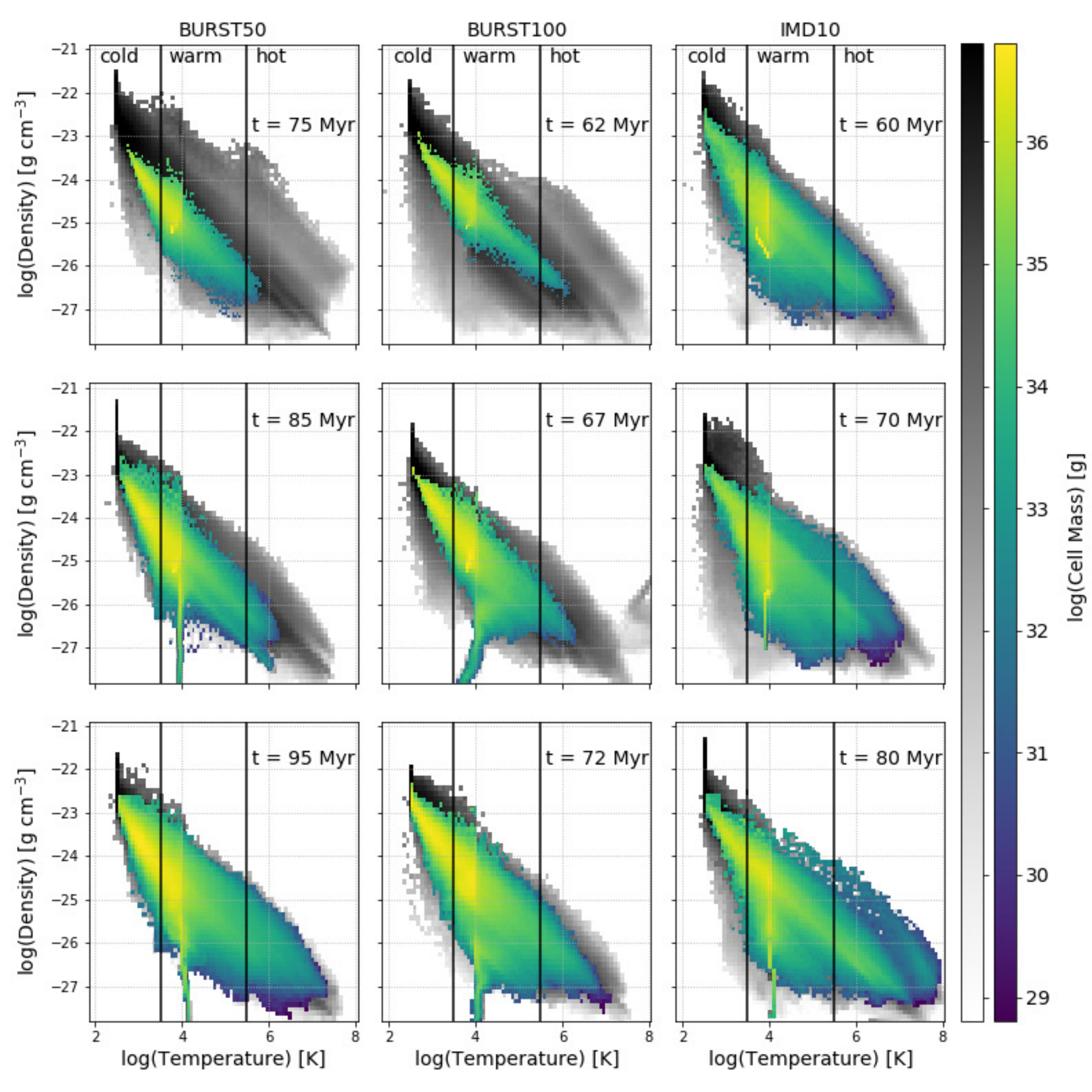}
\caption{Phase diagrams for the BURST50 simulation (left), BURST100 simulation (middle) and IMD10 simulation (right). These are 2D profiles of the average cell mass in each density-temperature bin. The grey scale maps the distribution of all of the gas in the simulation box. The injected gas distribution is overlaid with a colour scale (rightmost colour bar). We separate injected gas from the total gas content using the normalized colour field, which represents the fraction of injected gas present in a cell volume. Injected gas is selected using a normalized colour cut of $>50\%$. To demonstrate the gas phase evolution, we have plotted each simulation at three different time steps noted in the upper right corner of each panel, with earlier time steps at the top and later time steps at the bottom. Black vertical lines separate each panel into three temperature regimes: cold (T $\leq 10^{3.5}$K), warm ($10^{3.5}$K $<$ T $ < 10^{5.5}$K), and hot (T $\geq 10^{5.5}$K). }
\label{fig:phase}
\end{figure*}

The details of the global gas evolution are explored in Figures~\ref{fig:panels}$-$\ref{fig:phase}. Figure~\ref{fig:panels} shows slices in density, temperature, pressure, z-velocity, colour density and normalized colour density for four different time steps throughout the BURST50 simulation. At the boundary where the injected gas impacts the outflow, the injected gas decelerates rapidly and the z-velocity slices show a layer of stalled material. This corresponds to an over-dense layer in the injected gas density exceeding the initial conditions. In the temperature slices, this density enhancement allows the gas layer to cool efficiently, driving the temperature to $10^{4}$K and below. Rayleigh-Taylor fingers of cold, over-dense injected gas are seen falling into the less-dense halo environment at all stages of the time evolution in Figure~\ref{fig:panels}. These inflowing fingers fill low-density voids in the inhomogeneous outflow where there is only halo gas and the resulting shear drives Kelvin-Helmholtz instabilities with the adjacent outflowing gas, resulting in smaller fragments and generating strands of small clouds. The colour density and normalized colour density (lower two sets of panels) track the progress of the inflowing cloud as it fragments and mixes with the outflowing gas. We discuss the properties of these fragmented cloud structures in more detail in Section~\ref{sub:clouds}. 

The density and temperature slices in Figure \ref{fig:panels} show the multiphase structure generated by this interaction. The corresponding pressure slices generally show less variation on small scales (although by no means none: a rich set of shocks and other hydrodynamic features are clearly present). Instead, the (thermal) pressure variation is dominated by the mean profile as a function of height above the disk. We briefly examined how well the conditions of hydrostatic pressure equilibrium are fulfilled in these simulations, finding that the thermal pressure is, on average, sufficiently large to support the hot gas (see also Figure~\ref{fig:initial}); however, significant local pressure variations are common, in agreement with previous work \citep[e.g.,][]{Joung2006}. 

By the third panel (95 Myr) in Figure~\ref{fig:panels}, the velocity gradient across the kpc-long cloud has caused it to collapse into a relatively narrow high-density region, although with long cloud chains extending down toward the disk. Outflow begins to re-establish by the last time step (105 Myr), and will blow back a small portion of injected gas that did not manage to reach the disk (see also Section~\ref{sub:fate}). The pressure, temperature and velocity slices at this point show the first SN shocks behind the inflowing cloud. Movies showing the full frame-by-frame time evolution for all three simulations are provided in the supplementary materials. 

Figure~\ref{fig:profiles} shows one-dimensional profiles of the average density, temperature, pressure, and z-velocity as a function of vertical height in each simulation (including a run without inflow). The lighter transparent lines separate the properties of the injected gas (only) from those of the total gas using a normalized colour density cut of $>50\%$. For consistency, these profiles are made when the gas injection is complete at the last time step before the top boundary is switched back to outflow (i.e. the one-kpc cloud has just fully entered the grid). For the BURST50 simulation, this time step corresponds to the first panel (75 Myr) in Figure~\ref{fig:panels}. 

In all three simulations, as can be seen in the velocity profiles, the injected gas enters the box and accelerates (gravity only) towards the disk. It quickly decelerates when it reaches $z \sim 1.5-2.0$ kpc, forming a high-density reverse shock. This high-density region is spread between two drops in the velocity profile, beginning with a drop in the injected gas velocity and ending with a drop in the total gas velocity at the boundary where the injected gas and the outflow collide. The vertical extent and amplitude of the density enhancement in the region between the reverse shock and contact discontinuity are inversely related and set by the velocity gradient across the contact discontinuity, which varies between simulations. The 100 km s$^{-1}$ (BURST100) run experiences the sharpest deceleration upon impacting the outflowing gas, resulting in the highest density enhancement over a narrow ($\sim 0.25$ kpc) extent. The velocity gradient in the 10 km s$^{-1}$ (IMD10) run is much smaller and its broad, low-density reverse shock region spans $\sim 1$ kpc. In all simulations, injected gas that makes its way past the contact discontinuity through low-density gaps in the outflow resumes accelerating towards the disk and fragments into smaller clouds as seen in Figure~\ref{fig:panels}. These clouds are denser and cooler than the average surrounding medium. We discuss these cloud structures and the role of mixing in Section \ref{sub:clouds}.

In Figure~\ref{fig:phase} we show phase diagrams for each simulation at three different time steps. Again, for consistency, we compare across simulations at time steps with similar injected cloud heights. We have superimposed the injected gas properties (colour scale) on top of the total gas properties (gray scale) using the normalized colour density cut. Gas is initially injected in the warm phase with a temperature of $10^{4}$ K and a density of $\rm 8.37 \times 10^{-26} \ g \ cm^{-3}$. Over time the inflow distribution broadens to encompass a wide range of densities and temperatures; however, the majority of the injected gas mass cools to lower temperatures (below $10^4$ K) and higher densities.

\subsection{Cloud Properties} \label{sub:clouds}

\begin{figure*}
\centering
\includegraphics[width=\textwidth]{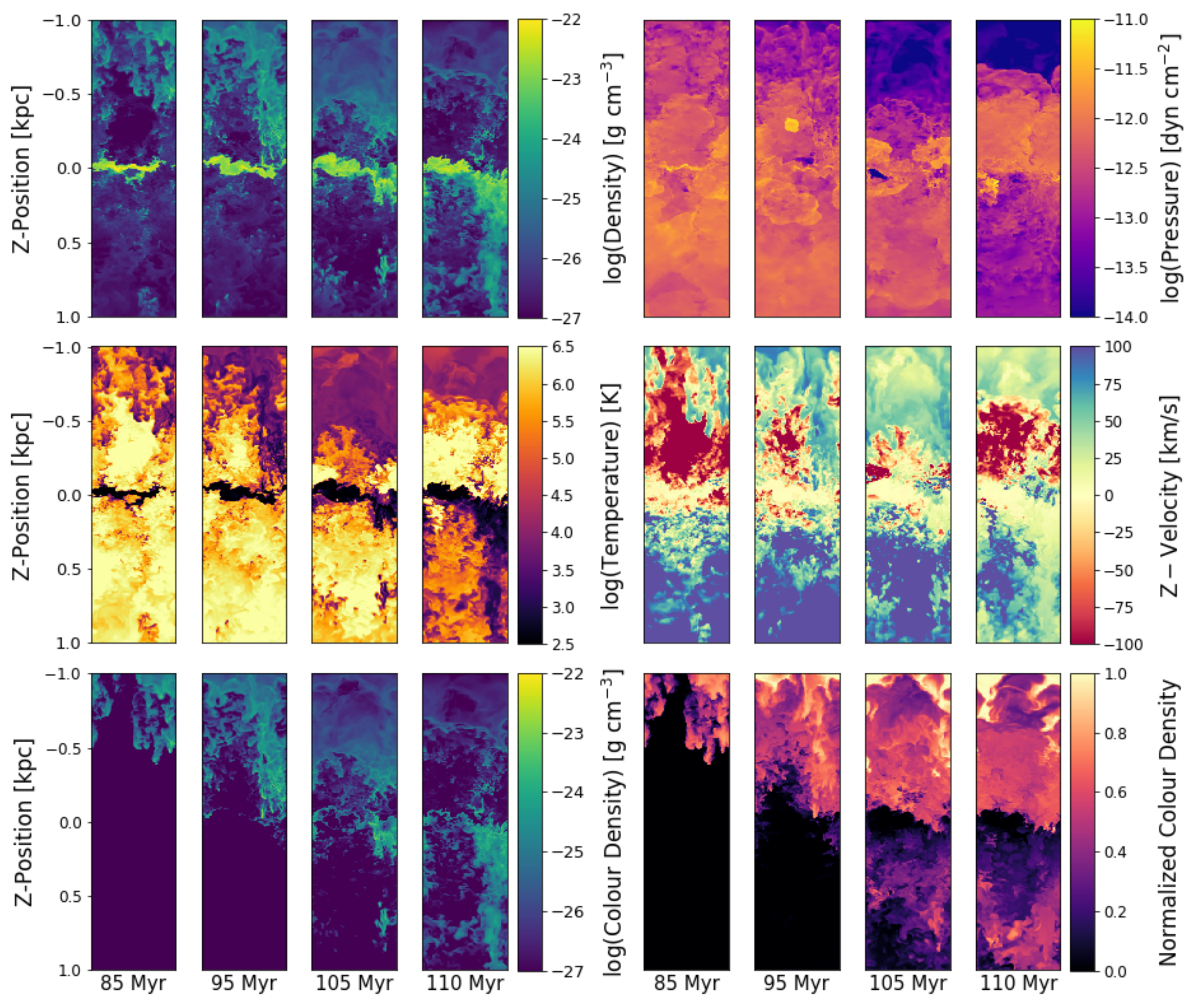}
\caption{Zoomed-in density, pressure, temperature, z-velocity, colour density, and normalized colour density slices showing cloud formation in the BURST50 simulation. Each panel is a slice in the y-z plane at x = 0 with dimensions $[0.35 \times 2]$ kpc. Note that z-axis is vertical and each slice encompasses the first region of 2x resolution refinement ($\pm 1$ kpc) and the second region of 4x resolution refinement ($\pm 0.5$ kpc). The plane of the disk is the yellow high-density region at z = 0. The bottom left panels show the colour density field, which traces the density of only the injected gas. The rest of the gas in the domain starts with a near-zero colour density field. We divide the colour density by the total gas density to create a normalized colour density field (bottom right), which represents the fraction of injected gas present in a cell volume. The time step for each panel is noted along the bottom of the plot. We have changed the aspect ratio in these slices (compressing them along the z-direction) to ensure they fit in this figure. } 
\label{fig:zoom}
\end{figure*}

In studying the gas evolution, we find that the injected gas fragments into strands of dense cold cloud structures (Figure~\ref{fig:panels}). In order to study these structures in detail, we create zoomed-in slices of the density, temperature, pressure, z-velocity, colour density, and normalized colour density surrounding the disk. These slices are shown in Figure~\ref{fig:zoom} for the BURST50 simulation. Each slice is centered on the galactic disk with a z-axis extent of $2$ kpc, encompassing the first resolution refinement at $\pm 1$ kpc and the second refinement at $\pm 0.5$ kpc. We have shifted the time step in each panel forward by 10 Myr relative to Figure \ref{fig:panels} in order to allow the injected gas time to extend into this region. The cleanest example of these structures is the second time step (95 Myr) in Figure~\ref{fig:zoom}, right before the injected gas hits the disk. The density, temperature and velocity slices show that these cloud structures are dense, cold and inflowing towards the disk. We see from the colour density structure that these clouds are composed of injected gas; however, in the normalized colour density structure, it is clear they are not pristine. Recall that the normalized colour density field tracks the fraction of inflowing gas in each cell (so that entirely pristine injected gas has a normalized colour density of 1). The injected cloud structure seen at 95 Myr has a normalized colour density of about [0.6 - 0.8], meaning that the injected gas in these cloud structures has mixed with the outflow/halo gas. The gas surrounding these dense cloud structures has a slightly lower normalized colour density but has also mixed with some of the injected gas. This mixing picture is consistent with the late time phase diagrams in Figure \ref{fig:phase}, where we see a growing component with higher temperatures and lower densities in the injected gas. With a normalized colour density cut of $50 \%$ we are certainly selecting some of this injected gas mixed with the hot, low density outflow. 

\begin{figure*}
\centering
\includegraphics[width=\textwidth]{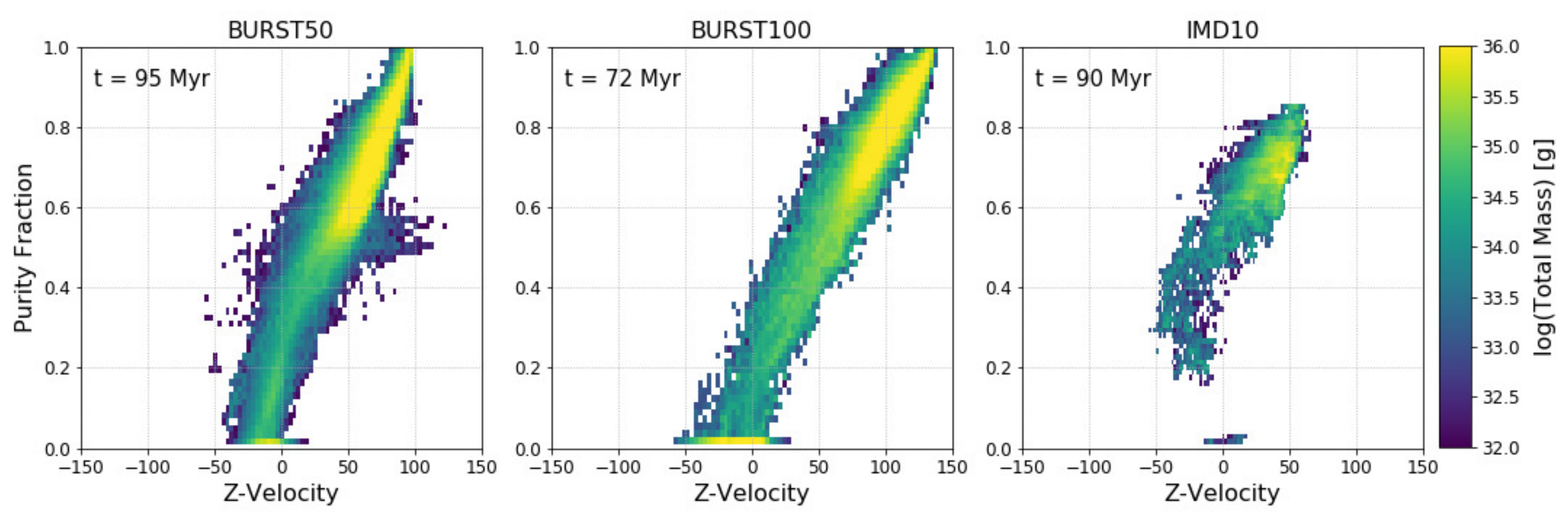}
\caption{In this plot we have isolated the properties of the cloud population above the disk using a z-position cut of z = $-0.2$ to z = $-1$ kpc and a density cut of $\rm 5 \times 10^{-25} \ g\ cm^{-3}$. For gas that meets this selection criteria, we generate a 2D distribution of the integrated mass in each purity$-$velocity bin. The purity fraction (also called normalized colour density) indicates the amount of mixing: a high value corresponds to pure inflow material. It is calculated as the ratio of the injected gas density in each cell (traced by the colour density field) to the total gas density in each cell. The velocity axis is in units of km $\rm s^{-1}$. We compare the three simulation runs: BURST50 (left), BURST100 (middle) and IMD10 (right). The time steps (noted in the upper left corner) are selected such that the injected gas in each simulation has just reached the galactic disk. For the BURST50 simulation, the selected time step matches the second panel in Figure \ref{fig:zoom}.} 
\label{fig:clouds}
\end{figure*}

In Figure~\ref{fig:clouds} we make a quantitative measurement of the mixing between the injected and outflowing gas across simulations. We have applied a z-position cut from z = -0.2 to z = -1 to select gas in the first refinement region above the disk (recall that negative z-position values are above the disk). An additional density cut ($\rm \rho > 5\times10^{-25} \ g \ cm^{-3}$) is applied to each simulation in order to select for gas in the observed cloud structures. 
%YZ: why you pick this density value instead of others to select the clouds? Does it enhance the cloud structures/masses or sizes, etc? 
Using this cut data, we plot the total mass distribution binned by purity fraction (normalized colour density) and z-velocity. Note that we have not made a purity cut to select only the injected gas; therefore, this plot includes the mass in high density outflow structures as well. In the BURST50 and BURST100 simulations, we find that the majority of the sampled gas lies in two components: a very low-purity outflowing component (with z-velocity $<$ 0) and a higher-purity inflowing, mixed component. Most of this higher-purity component is not pristine, and extends down to a purity fraction of $50\%$ in the BURST50 simulation. We find slightly less mixing in the BURST100 simulation, where the majority of the gas mass does not mix below $\sim 60\%$. In the IMD10 run we only see one component. Both the low-purity outflowing gas and the pristine gas are completely absent and all of the selected gas is mixed with a purity fraction of $\sim 40 - 90 \%$. Note the strong correlation between inflow velocity and purity fraction, an indication that deceleration of the clumps is driven primarily by mixing with the outflowing gas. See also the lower panel in Figure \ref{fig:initial}; the pressure of the outflowing gas cannot contribute much to the deceleration of the inflow, suggesting an  alternate source such as mixing is responsible. 

\begin{figure*}
\centering
\includegraphics[width=\textwidth]{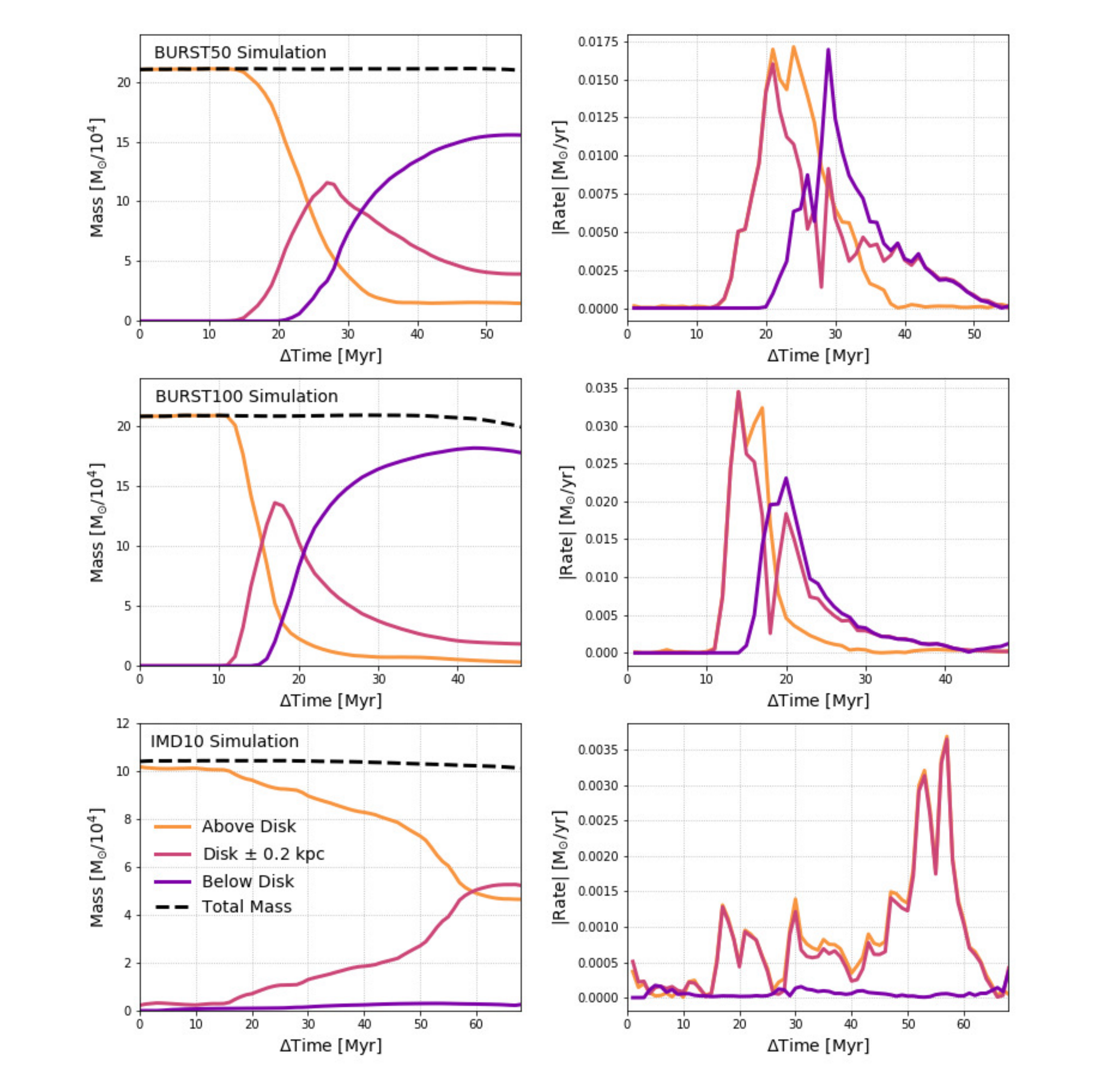}
\caption{In this figure, we split the simulation box into three vertical segments and follow the flow of injected gas through the segments over time. We define a segment below the disk extending from $z = 0.2$ kpc to the bottom boundary of the box at $z = 2.5$ kpc and a complementary segment above the disk extending from $z = - 0.2$ kpc to the top z-boundary edge of the box at $z = - 2.5$ kpc. A middle segment centered on the galactic disk extends from $- 0.2$ to $+ 0.2$ kpc. We track the injected gas through the simulation box using the colour density field, which is the injected gas density in a given cell volume. (Left) The integrated injected gas mass in each of three simulation box segments over time. The injected mass in each segment is calculated by summing the product of the colour density and the cell volume. The dashed black line is the total injected mass (See Table \ref{table:parameters}). (Right) The absolute value of the mass injection rate in each of the three segments. All plots are shown as a function of time since the injection start time (See Table \ref{table:parameters}). The top, middle, and bottom panels are for the BURST50, BURST100, and IMD10 simulations respectively.} 
\label{fig:fate}
\end{figure*}

\begin{figure}
\centering
\includegraphics[width=0.87\columnwidth]{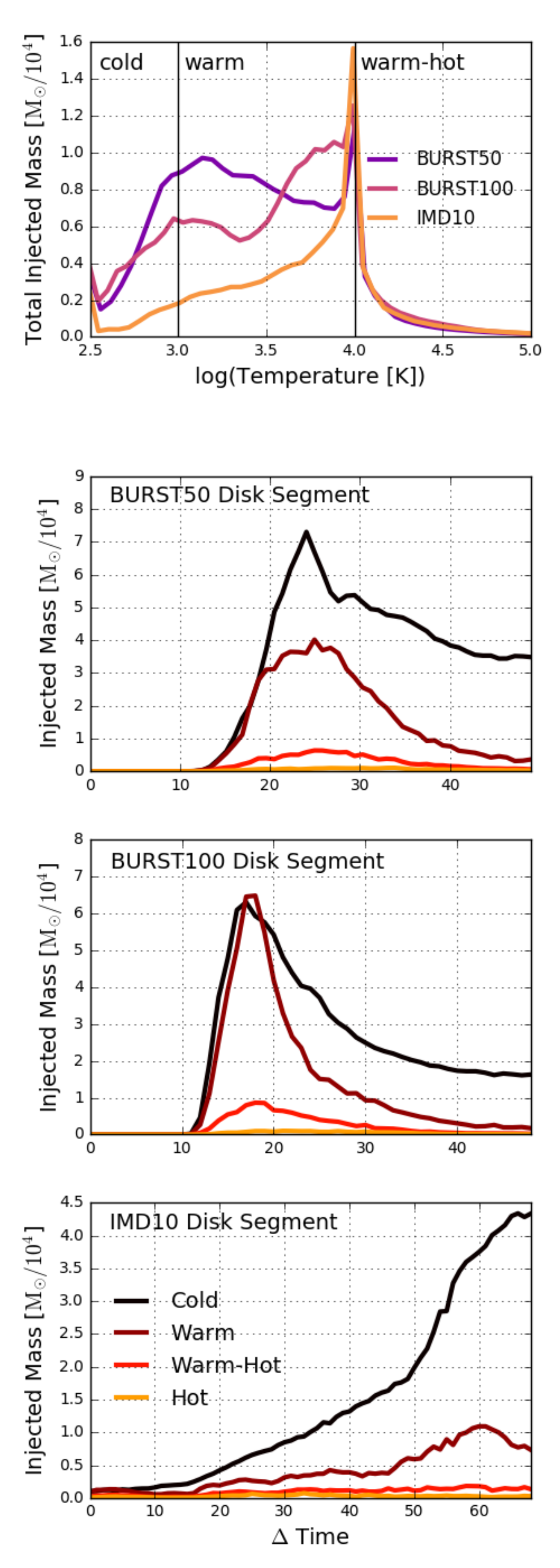}
\caption{The top panel shows the average mass as a function of temperature for the injected gas throughout each simulation. These profile were made at the time step directly before the gas hits the disk. The bottom three panels focus only on injected gas in the disk segment ($|z| < 0.2$ kpc). The mass of injected gas inside the disk segment was calculated the same way as in Figure \ref{fig:fate}, but here we additionally split the injected gas into four temperature bins. Temperature cuts are defined as: cold (T $< 10^{3}$ K), warm ($\rm 10^{3} \leq T < 10^{4}$ K), warm-hot ($\rm 10^{4} \leq T < 3 \times 10^{5}$ K), and hot ($ \rm T \geq 3 \times 10^{5}$ K)}. 
\label{fig:temp}
\end{figure}

\subsection{Inflow Fate} \label{sub:fate}
Injected gas that makes it to the galactic disk will either successfully accrete onto the disk or pass through it entirely. 
We study the fate of the injected gas by splitting the simulation box into three segments. Each segment is a smaller box with the same x-y dimensions as the simulation box and a non-overlapping vertical extent. The middle segment is centered on the galactic disk, extending $\pm 0.2$ kpc in the z-direction. The other two segments are above and below the disk starting at z = $\pm 0.2$ kpc and extending to the edge of the box at z = $\pm 2.5$ kpc. These three segments combined encompass all of the gas in the simulation. We calculate the mass of injected gas in each segment by integrating the product of the colour density and the cell volume throughout each segment. We do not use the normalized colour density cut, as discussed in previous plots, because this quantity includes some of the outflow mixed with the injected gas and would violate our expectation of injected mass conservation as material is exchanged between segments.

Figure \ref{fig:fate} shows the integrated mass of the injected gas (left) and the mass injection rate (right) in each segment as a function of time. Note that the total injected gas mass (dashed line) is a roughly constant value equal to the total mass in Table \ref{table:parameters}. The only time the injected mass is lost from the system is towards the end of the simulation when it either passes through the disk and exits the bottom z-boundary or gets swept up in the re-established outflow and exits the top z-boundary. At the start of each simulation, a finite gas mass has been injected into the top segment above the disk. As the injected gas moves towards the disk it begins spilling over into the disk segment. Any mass that does not stick in the disk then passes through to the bottom segment below the disk. At the finial time step in Figure \ref{fig:fate}, the injected gas mass in the disk segment has been accreted. Mass remaining above or below the disk has been pushed back towards the top boundary or passed through the disk respectively. We find in Figure~\ref{fig:fate} that the fate of the injected gas depends strongly on the injection velocity. In the faster 50 km s$^{-1}$ and 100 km s$^{-1}$ runs, nearly all of the gas makes it to the disk, but only a small fraction (20$\%$ of the total injected gas mass in the BURST50 simulation and 10$\%$ in the BURST100 simulation) sticks in the disk and actually ``accretes". The majority of the injected gas in these high-velocity runs plunges straight through the disk into the bottom segment. This picture is confirmed in the late-stage time evolution panels in Figure \ref{fig:zoom}, where large clouds of injected gas are seen exiting the disk and outflowing towards the bottom z-boundary. In the 10 km s$^{-1}$ run, only about half of the injected gas makes it to the disk, but nearly all of that mass (50$\%$ of the injected gas mass) accretes. The mass that does not make it into the disk segment in the IMD10 run is swept up in the re-established outflow and exits the top z-boundary of the box. 

In Figure~\ref{fig:temp} we quantify the temperature of the injected gas accreting onto the galactic disk. The top panel shows the initial temperature distribution of the injected gas in each simulation at a time step directly before the gas hits the disk (panel 3 in Figure \ref{fig:panels}). We see a spike in all simulations at the injection temperature of $10^{4}$ K. However, in all three simulations the gas quickly redistributes towards lower temperatures. The bottom three segments show the injected gas temperature evolution as a function of time in the disk segment of each simulation. Here we have split the gas into four temperature regimes: cold gas (T $< 10^{3}$ K), warm gas ($\rm 10^{3} \leq T < 10^{4}$ K), warm-hot gas ($\rm 10^{4} \leq T < 3 \times 10^{5}$ K), and hot gas (T $\geq 3 \times 10^{5}$ K). While a large fraction of warm gas does pass through the disk in the BURST50 and BURST100 simulations, the large majority of the accreted injected gas present in the disk at the end of all three simulation has cooled below $10^{3}$ K. 

% -----------------------------------------
\section{Observational Predictions} \label{sec:obs}
We use this suite of simulations to predict observational signatures of gas accretion. These predictions were made using reference tables generated by \cite{2016ApJ...827..148C} with the photoionization code \texttt{CLOUDY} \citep{1998PASP..110..761F}. Specifically, we are using the g1q1 run at redshift z = 0 detailed in \cite{2016ApJ...827..148C}. This model was interpolated over a temperature range spanning $[10^{3} - 10^{8}]$ K and a number density range spanning $[10^{-6} - 10^{2}] \rm \ cm ^{-3}$. Density and temperature values from our simulation suite were referenced with these interpolated \texttt{CLOUDY} tables to model the ion fraction and emissivity used in the following analysis. In Section \ref{sec:density} we generate mock column density observations in HI and OVI. We chose these two ions because they have observable counterparts and are representative of low and high-energy ionization species. Simiarly, we create surface brightness maps of H$\alpha$ and OVI emission in Section \ref{sec:emission}. The photoionzation model applied assumes an extragalactic UV background (EUVB) as well as solar metallicity and abundances. This model does not include a source of ionizing photons from the disk. We do not expect ionizing radiation from the disk to significantly affect our OVI predictions, as we find collisional ionization is the dominant source of OVI ionization in our simulations. Ionizing photons from the disk may significatly affect the the HI and H$\alpha$ results; however, accurately accounting for radiative transfer of hydrogen ionizing radiation is a beyond the scope of this paper \cite[e.g.,][]{Wood2010}.

\subsection{Absorption: Column Densities of HI and OVI} \label{sec:density}
In order to calculate the number density $n_{Xi}$ for an element $X$ in ionization state i (e.g. HII: $X =$ H and $i = 2$), we substitute the simulation density ($n_{H}$) and the ion fraction computed by \texttt{CLOUDY} $(n_{Xi}/n_{X})$ into Equation \ref{eq:colN}. For metal ions, we add a factor for the elemental abundance ($n_{X}/n_{H}$) relative to hydrogen, which is the simulated metallicity scaled by the solar abundance \citep{2016ApJ...827..148C}. We create an approximate metallicity field, assuming that the injected gas has a metallicity of  $0.2Z_{\odot}$ \citep{2017ApJ...837..169P}. The rest of the gas in the simulation volume is given solar metallicity, including the disk, outflow, and disk-halo gas.

\begin{equation}
n_{Xi} = n_{H}(n_{X}/n_{H})(n_{Xi}/n_{X}) \ \ \rm [cm^{-3}]
\label{eq:colN}
\end{equation}

\begin{figure*}
\centering
\includegraphics[width=\textwidth]{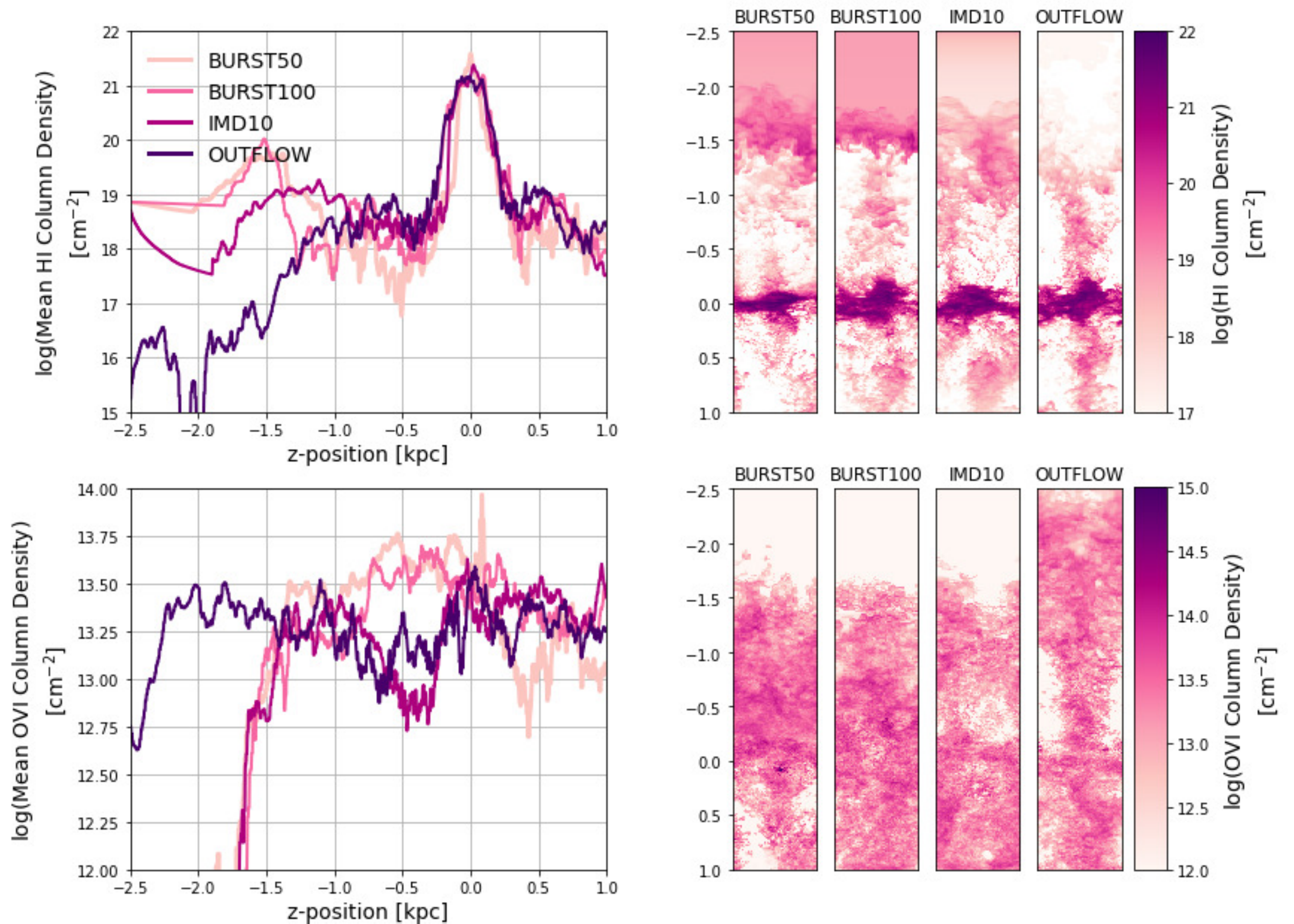}
\caption{Edge-on column density predictions for HI (top) and OVI (bottom) compared at time steps where the injected cloud has reached $\sim$ 1 kpc in the BURST50, BURST100, and IMD10 simulations. The right panels are 2D maps of the predicted column density projected along the x-axis (integrated over the 0.35 kpc simulation width). The left panels are 1D profiles of the average column density binned by z-position. A negative z-position is above the disk, and a positive z-position is below the disk. These three simulations are compared with column density predictions for the final outflow time step before inflowing gas is injected} 
\label{fig:col}
\end{figure*}

Once we have the ion number density $n_{Xi}$, the projection of $n_{Xi}$ along the line of sight (the x-axis of the simulation box) is the column density shown in Figure \ref{fig:col}. Here we compare the BURST50, BURST100, and IMD10 simulations. In addition, we have included an outflow panel (without any inflowing gas) at the time step immediately preceding the gas injection. The right panels in Figure \ref{fig:col} are 2D edge-on column density maps of HI (top) and OVI (bottom). In the left panels, we have averaged across the width of each projection, creating 1D profiles of column density with respect to z-position. These maps reveal a number of interesting structures in the injected gas, outflowing gas, and galactic disk. We compare these predicted column density features with observations. We restrict our comparison to observations of the Milky Way because the initial conditions in this simulation were modeled after the solar neighborhood. In addition, the proximity of studying our own galaxy affords us a wealth of data.

Regions of high HI column density seen in the top panels of Figure~\ref{fig:col} probe cold neutral gas. This gas is especially prominent at three locations: (i) the galactic disk, (ii) behind the shock front, and (iii) in the cold high-density cloud structures above the disk. We discuss each of these features in comparison with observations of atomic hydrogen 21-cm emission in the Milky Way. 

In the galactic plane, measurements of the HI distribution probe the structure of the galactic disk and the properties of the interstellar medium. From Figure~\ref{fig:col}, we find a mean log HI column density of $[21.1 - 21.5]$ at the disk mid-plane across the simulations. We compare this prediction with maps of the integrated HI 21-cm emission intensity from the Leiden-Argentine-Bonn (LAB) all-sky survey \citep{2005A&A...440..775K}. Using the LAB data cube, the mean log HI column density integrated over the velocity range $\pm 400$ km s$^{-1}$ within $\pm 10^{\circ}$ of the Galactic plane is 21.5. This number is consistent with our predictions; however it may be a lower limit on the true Galactic column density because it was derived under the assumption that HI 21 cm emission in the Galactic plane is still optically thin. 

Beyond the Galactic plane, a number of HI features are observed at the disk-halo interface. In our simulations we find dense cloud structures within about 1 kpc of the disk. As seen in the outflow panel in Figure~\ref{fig:col}, some of these cloud structures are associated with the SN feedback. The rest are formed from the fragmented injected gas inflowing towards the disk. In Figure~\ref{fig:col}, both sets of clouds have HI column densities of a few times $\rm 10^{19} \ cm^{-2}$. Similar neutral hydrogen clouds have been observed at the disk-halo interface of the Milky Way in a number of HI surveys. Tangent point observations of 21 cm emission with the Green Bank Telescope (GBT) discovered $\sim 30$ such cloud structures within 0.5 to 1.5 kpc of the disk with HI column densities in the range [0.4 - 2.7] $\rm \times 10^{19} \ cm^{-2}$ \citep{Lockman2002}. These clouds are co-rotating with the galactic disk, exhibiting only low-velocity deviations on the order of 15 km s$^{-1}$. Recent advancements in observational survey sensitivity, speed, and resolution have led to the cataloging of thousands of neutral hydrogen clouds surrounding the Milky Way. The GALFA-HI survey conducted at the Arecibo Observatory identified 1245 low to intermediate velocity clouds with HI column densities ranging from $[0.1-2.8] \rm \times 10^{19} \ cm^{-2}$ \citep{Peek2011,Saul2012}. This sample is divided into cold and warm clouds, with median column densities of $\rm 0.6 \times 10^{19} \ cm^{-2}$ and $\rm 0.9 \times 10^{19} \ cm^{-2}$ respectively. The cloud column densities in our simulations are certainly consistent with the broad range of low-intermediate velocity cloud properties, but they exceed the mean observational column densities. Its possible that this column density excess is caused by the 2 pc minimum grid resolution in our simulations, which constrains clouds from fragmenting below the resolution limit. 

At slightly higher latitudes, our simulations predict significant HI column densities behind the shock front in the pile-up of injected gas. This wall feature is produced by the sudden deceleration of the injected gas in contact with the SN-driven outflow. Gas in this density enhancement is able to cool down to (and below) $10^{4}$ K, increasing the neutral fraction and raising the HI column density to $\sim 10^{19} - 10^{20} \rm \ cm^{-2}$, as seen in Figure~\ref{fig:col}. This HI feature is much less pronounced in the IMD10 simulation due to the smaller velocity gradient across the injected gas cloud (and resulting lower densities). Similar high-density wall/ridge structures have been observed alongside high-velocity clouds interacting with the Galactic halo and ISM. We compare with the Smith Cloud, a high-velocity cloud interacting with the Galactic halo. The Smith Cloud is rather large, with an area of $\sim 3 \times 1$ kpc and an HI mass of $ \geq 10^{6}  \rm \ M_{\odot}$; however, the tip of the Smith Cloud is only 2.9 kpc below the galactic plane, making it a rough analogue to the injected gas in our simulations. GBT 21-cm observations of the Smith Cloud show a decelerated ridge of high-density gas $\rm N_{HI} = 2 \times 10^{20} \ cm^{2}$ along the edge of the cloud \citep{Lockman2016, Lockman2008}. The large HI mass of this ridge associates it with decelerated gas from the Smith Cloud and not swept-up material from the surrounding halo gas \citep{Hill2013}. The Smith Cloud also appears to be fragmenting at this boundary, exhibiting indications of Rayleigh Taylor and Kelvin Helmholtz instabilities as it interacts with gas in the Galactic halo \citep{2015IAUGA..2257086L}

In contrast to HI, regions of high OVI column density probe highly ionized, usually hot ($10^{5.5}$ K) gas. We find that collisional ionization is the primary ionization source for the OVI gas in our models \citep[see also][]{Smith2011}. Accordingly, the distribution of OVI in our simulations primarily traces the supernovae driven outflow. We compare our OVI column density predictions with \cite{Li2017b}, who performed a detailed study of the OVI density distribution in simulated outflow generated using an identical SN prescription. In \cite{Li2017b}, the OVI density was calculated using the product of the O/H solar abundance ($5\times 10^{-4}$) and the temperature-dependent OVI ionization fraction from \cite{Sutherland1993}. The OVI column densities stated in \cite{Li2017b} were calculated for a time-step of 140 Myr with error bars noting the time variation in the latter half of the simulation. Our outflow time step of 50 Myr is therefore outside the range of these error bars. Although we cannot make a direct comparison between the two sets of the predictions, we note similarities and differences below. Using CLOUDY, we predict an average OVI density of  $n_{OVI} = 1.8 \times 10^{-8} \rm \ cm^{-3}$ within $\pm 100$ pc of the disk. This is roughly consistent (to within $2\sigma$) with the average OVI density of $n_{OVI} = (2.4 \pm 0.3)\times 10^{-8} \rm \ cm^{-3}$ predicted by \cite{Li2017b} within $\pm 100$ pc of the disk. When we calculate the average OVI density within $\pm 200 $ pc from the disk at 50 Myr we find that the value remains approximately constant at $1.8 \times 10^{-8} \rm \ cm^{-3}$. \cite{Li2017b} reports a very similar average OVI density of $n_{OVI} = (1.8 \pm 0.3) \times 10^{-8} \rm \ cm^{-3}$ within $\pm 200$ pc at 140 Myr, which has dropped relative to their prediction close to the disk. All of these values are consistent with absorption line studies of nearby stars, which measure the OVI density of the local ISM to be $n_{OVI} = (2.2 \pm 1.0) \times 10^{-8} \rm cm^{-3}$ \citep{Jenkins1978b,Oegerle2005,Savage2006,Bowen2008,Barstow2010}.

\subsection{Emission} \label{sec:emission}
\begin{figure*}
\centering
\includegraphics[width=\textwidth]{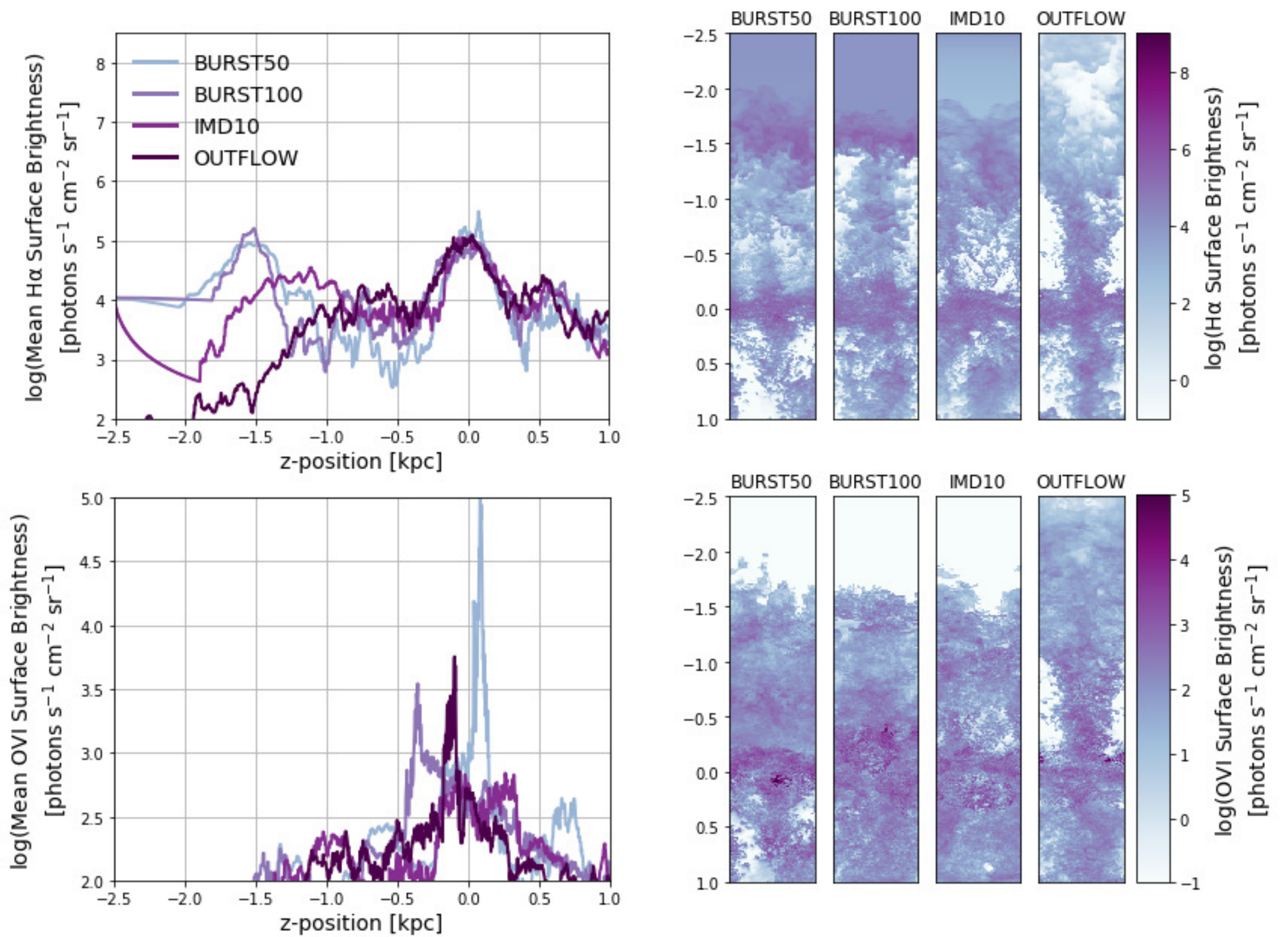}
\caption{Edge-on emission surface brightness predictions for H$\alpha$ (top) OVI (bottom) compared at time steps where the injected cloud has reached $\sim$ 1 kpc in the BURST50, BURST100, and IMD10 simulations. The right panels are 2D maps of the predicted surface brightness projected along the x-axis (integrated over the 0.35 kpc simulation width). The left panels are 1D profiles of the average surface brightness binned by z-position. A negative z-position is above the disk, and a positive z-position is below the disk. These three simulations are compared with the emission predictions for the final outflow time step before inflowing gas is injected.} 
\label{fig:emis}
\end{figure*}

We next turn to line emission as a complementary probe of the disk-halo conditions, again for the same two ions. The emission intensity ($\Gamma$) is calculated from the \texttt{CLOUDY} emissivity using Equation~\ref{eq:emis}. Here $\epsilon(\rho, T)$ is the temperature and density dependent emissivity, $h$ is Planck's constant, $c$ is the speed of light, and $\lambda$ is the wavelength of the recombination transition ($\lambda 6563 \AA$ for $H\alpha$ and $\lambda 1032 \AA$ for OVI). The projection of the emission intensity integrated along the line of sight is the surface brightness shown in Figure~\ref{fig:emis}. On the right we have plotted edge-on projections of the $\rm H\alpha$ and OVI emission surface brightness for each of the three simulations and compared them with surface brightness predictions for an outflow time step preceding the gas injection. The corresponding average emission profiles as a function of z-position are plotted on the left panels. 
\begin{equation}
\Gamma = \frac{\epsilon(\rho, T)}{4\pi} \frac{\lambda}{hc} \rm \ \ [photons \ s^{-1} \ cm^{-2} \ sr^{-1}]
\label{eq:emis} 
\end{equation}

The H$\alpha$ emission in Figure~\ref{fig:emis} traces high-density structures of warm, 
%YZ: the cold/warm terms are still confusing, so may as well put the temperature range in () to remind the reader. 
partially ionized gas and highlights regions of strong gas cooling. Most notably, regions of high H$\alpha$ emission intensity correspond to (i) the high-density ISM ejected from the disk via supernovae feedback, and (ii) the dense wall of injected gas piling up behind the shock front. 

We compare these predictions with observations from the Wisconsin H$\alpha$ Mapper (WHAM) \citep{Haffner2003}. WHAM surveyed the distribution and kinematics of H$\alpha$ emission in and around the Milky Way. These observations probe the warm/warm-hot gas in the ISM as well as revealing a number of H$\alpha$ structures at the disk-halo interface. In Figure~\ref{fig:emis} the galactic disk and supernova-driven outflows produce H$\alpha$ emission intensities ranging from $\sim 10^{3}$ to $10^{5} \rm \ photons \ s^{-1} \ cm^{-2} \ sr^{-1}$. Gaps in the inhomogeneous outflow reveal the hot halo gas with essentially no H$\alpha$ emission intensities. As with the column density, we expect to under-predict the emission intensity due to the shortened projection path length (0.35 kpc) through our simulation box. WHAM observations along the Sagitarius-Carina Arm reveal similar H$\alpha$ emission values ranging from $> 2 \times 10^{6} \rm \ photons \ s^{-1} \ cm^{-2} \ sr^{-1}$ in the galactic plane to $\sim 1 \times 10^{4} \rm \ photons \ s^{-1} \ cm^{-2} \ sr^{-1}$ beyond the disk \citep{2017ApJ...838...43K}. Models of the disk scale-height place these emission measurements within a vertical height of 3 kpc of the Galactic disk, making them an excellent analog for our simulation box height. The WHAM emission maps also reveal patches below the detection limits similar to the predicted gaps in our H$\alpha$ emission maps. 

Another prominent H$\alpha$ feature in our simulation suite is the dense wall of decelerated injected gas. In Figure \ref{fig:emis} we see a thick wall of elevated H$\alpha$ emission trailing the leading edge of the injected gas. Looking closely, this wall is actually composed of many thin density enhancements projected on top of one another in the edge-on inclination. This effect is due to the inhomogeneous outflow halting the injected gas at different heights above the galactic disk. As mentioned in Section \ref{sec:density}, we compare this feature with the Smith Cloud, an HVC at the Milky Way disk-halo interface interacting with the ISM. More than half the mass of the Smith Cloud is ionized, producing a smooth distribution of H$\alpha$ emission \citep{Wakker2008, Hill2009}. In addition, H$\alpha$ observations with WHAM reveal a ridge of elevated H$\alpha$ emission trailing the HI ridge by a few hundred pc. Although the HI and H$\alpha$ ridges are not coincident, their similar morphology suggests a physical connection. Gas in this ionized feature is likely shock heated, with an H$\alpha$ emission intensity ranging from $[3.17-5.56]\times10^{4} \rm \ ph \ s^{-1} \ cm^{-2} \ sr^{-1}$ \citep{Hill2013}. In our simulations, H$\alpha$ emission associated with the deceleration of injected gas ranges in mean intensity from about $[1 - 15] \times10^{4} \rm  \ ph \ s^{-1} \ cm^{-2} \ sr^{-1}$ (Figure \ref{fig:emis}). Overlapping H$\alpha$ ridges are seen trailing the cloud front by as much as half a kpc. While we cannot make a direct comparison with the neutral and ionized ridge morphology observed in the Smith Cloud, 
%YZ: be a bit more specific about why you can't make the direct comparison? 
we note that we are able to reproduce a reasonable range of elevated H$\alpha$ emission intensities where the shocked ionized component is separated from the decelerated neutral gas. 

Similar to H$\alpha$, OVI emission also traces gas cooling, but in the highly ionized hot gas. In agreement with the low OVI column densities predicted in the injected gas in Figure \ref{fig:col}, we do not predict significant OVI emission from the injected gas in Figure \ref{fig:emis}. We conclude that the injected gas is not heated (above $10^{5.5}$ K) and re-cooled on its path the the disk. However, we do predict significant OVI emission from the SN-driven outflow with an average intensity of  $1.84\times 10^{4} \rm \ ph \ s^{-1} \ cm^{-2}  \ sr^{-1}$ projected along the z-axis. We compare our {\sc Cloudy} prediction with OVI emission predictions from \cite{Li2017b} which were calculated using the OVI ion fraction described in Section \ref{sec:density} and an oscillator strength of 5 for the $\lambda 1038 \AA$ emission line \citep{Osterbrock1989}. As noted in Section \ref{sec:density}, \cite{Li2017} makes emission predictions at a time step of 140 Myr with error bars time averaged over the later half of the simulation. Our final outflow time step at 50 Myr falls outside the range of these error bars; however, we still make an approximate comparison. The z-axis projection in \cite{Li2017b} predicts an OVI $\lambda 1032 \AA$ emission intensity of $[0.5 - 6] \times 10^{3} \rm \ ph \ s^{-1} \ cm^{-2} \ sr^{-1}$ at a time step of 140 Myr, which is about a factor of three lower than our emission predictions. Note that both our predictions and the \cite{Li2017b} predictions are the result of projecting through the entire simulation box (along the z-axis), and need to be divided by two in order to compare with the following observations taken within the Milky Way. Observations with FUSE detected OVI $\lambda 1032 \AA$ emission along 23 sight lines with intensities ranging from $(1.9 - 8.6) \times 10^{3} \rm \ ph \ s^{-1} \ cm^{-2} \ sr^{-1}$ \citep{Otte2006}. Despite extinction, higher emission intensities in this survey were actually found at low-latitude sight lines in the nearby ISM. In addition, high-latitude observations of OVI emission have been obtained with SPEAR close to the North Galactic Pole, measuring an emission intensity of $(0.5 - 2.0) \times 10^{4} \rm \ ph \ s^{-1} \ cm^{-2} \ sr^{-1}$ \citep{Welsh2007}. Our OVI emission predictions (as well as the prediction from \cite{Li2017b}) are roughly consistent with this range of observational values. 

% -----------------------------------------

\section{Discussion} \label{sec:dis}
In this work, our inflowing gas cloud was deliberately idealized: we assumed a constant density, temperature and inflow velocity. This was done in part because our goal is to understand the physical processes operating in the inflow/outflow interaction, and our outflow properties are already quite complicated, with multiphase gas ejected from realistically generated supernovae explosions. We briefly discuss physical origins for this inflowing gas cloud in Section \ref{subsec:origins} and elaborate on select results from Section \ref{sec:results} including the evolution of the injected cloud structures (Section \ref{subsec:cloudevo}) and the flow of injected gas plunging through the galactic disk (Section \ref{subsec:pastdisk}). We compare this work with the existing literature (Section \ref{subsec:compare}) and make note of some of the simulation limitations and assumptions (Section \ref{subsec:caveats})

\subsection{Inflow Origin}
\label{subsec:origins}
The original source of the injected gas cloud in this simulation could be either cosmological inflow, gas stripping from satellites, or gas recycling, all of which are expected to fuel the galactic disk. Cosmological inflow in the form of substantial cold filamentary flows has been seen in a wide variety of simulations \citep[e.g.,][]{Keres2009,Nelson2013,Hobbs2013,Joung2012}. The properties of these flows close to the disk-halo interface are not entirely clear. As demonstrated by the breadth of work discussed in Section \ref{subsec:compare}, understanding the flow of cold clouds through a hot CGM is challenging and requires additional numerical, observational and analytic studies. However, it seems possible and maybe even likely that at least some gas clouds with a cosmological origin can get close to the disk without being completely mixed into the CGM \citep{Stewart2017}.

Additionally, the inflow in these simulations could originate from recycled gas flow. Assuming that outflows are highly mass loaded (as seem to be required to reproduce galaxy properties), much of the ejected gas will fall back into the CGM, and eventually back onto the galactic disk. This was found in early (relatively low-resolution) simulations of momentum-driven winds \citep{Oppenheimer2010} but has since been confirmed and quantified in higher resolution simulations with more physically based feedback \citep{Muratov2015, AA2017}.

\subsection{Cloud Evolution}
\label{subsec:cloudevo}
Despite our simplified initial conditions, we found that the injected gas rapidly fragmented into small clumps through a series of instabilities, both hydrodynamic (Rayleigh-Taylor, Kelvin-Helmholtz) and thermal (via radiative cooling). These clumps initially accelerated ballistically from their starting inflow velocity (see Figure~\ref{fig:profiles}). The amount of acceleration immediately after injection (before slowing due to interactions with the outflowing material) was larger for simulation runs with a lower injection velocity (e.g. the IMD10 gas gained 40 km s$^{-1}$ while BURST100 only added 20 km s$^{-1}$ to its initial velocity).  

Over time, the inflowing clouds in all three simulation runs were decelerated as they interacted with the outflowing gas. As demonstrated in Figure~\ref{fig:clouds}, clouds within 1 kpc of the disk have a narrow velocity distribution that appears to peak around a typical cloud velocity in each simulation ($\sim$ 40 $\rm km \ s^{-1}$ in the IMD10 simulation to $\sim$ 100 km s$^{-1}$ in the BURST100 simulation). It is tempting to associate this with a ``terminal" velocity; however, the analogy with the classic computation should not be taken too far, as the clouds are not coherent entities. Unlike a penny dropped from a tall building, the clouds have no interstitial forces to hold them together. Instead, as evident from the tight correlation in Figure~\ref{fig:clouds}, deceleration is directly associated with gas mixing. Therefore, a simple momentum balance argument is probably a better explanation of the resulting cloud velocity distribution.

Irrespective of the physics causing the cloud deceleration, it appears that the fragmentation of our injected gas into dense ($n \sim 1$ cm$^{-3}$), cold ($T  < 10^4$ K) clumps with velocities of 50-100 km s$^{-1}$ is difficult to avoid. Even in our IMD10 case, where the inflow ram pressure is below the outflow ram pressure, we find inflowing clouds penetrating down to the disk with velocities of ~40 km s$^{-1}$. In this low-velocity case, fragmentation of the injected cloud results in dense cloud clumps that are able to overcome the outflowing ram pressure. Radiative cooling is key to the creation of these cold dense clouds and should be traced by the associated Ly$\alpha$ and H$\alpha$ emission (as described in Section \ref{sec:obs}).

Another interesting aspect of the cloud evolution is the spatial correlations in the clouds. This can be seen in Figure~\ref{fig:zoom} but is much clearer in the supplied movies. The clumps are not randomly distributed throughout the hotter background material filling the disk-halo interface, but tend to occur in chains or filaments (which themselves are clustered). This clustering appears to be due to a combination of effects. The physical process of cloud fragmentation causes larger clouds to  break up into smaller localized groups, and this group morphology facilitates a hydrodynamic shielding effect (like Canadian geese flying in formation) that prevents stripping and disruption.

\subsection{Outflow Past the Disk}
\label{subsec:pastdisk}
As mentioned in Section \ref{sub:fate}, some of the injected gas passes through the galactic disk and continues towards the bottom boundary of the simulation box. This is especially prevalent in the BURST50 and BURST100 simulations where the inflowing gas has higher ram pressure and is able to penetrate through low surface density regions in the disk. The bottom panels in Figure \ref{fig:zoom} show the inflow colour density and normalized colour density for gas within $\pm 1$ kpc from the galactic disk in the BURST50 simulation. Recall that the inflow colour density is the density of the injected gas and the normalized colour density is the ratio of injected gas density to total gas density. The third (105 Myr) and fourth (110 Myr) inflow colour density time steps in Figure \ref{fig:zoom} show the 50 km s$^{-1}$ injected gas plunging through the galactic disk. This gas is then swept away from the galactic plane with the SN outflow. Before hitting the disk, the injected gas has a high normalized colour density ($ > 0.6$). Some mixing with the outflowing gas has occurred, but the injected gas is still close to pristine. After exiting the disk, the injected gas exhibits a broader range of normalized colour density values. We find clumps of injected gas with high normalized colour density values which have not mixed significantly during their passage through the disk. However, we also find injected gas with low ($ < 0.6$) normalized inflow colour density values, suggesting that some of the injected gas passing through the galactic disk has mixed with the ISM. 

\subsection{Inflow Extent}
\begin{figure}[h!]
\centering
\includegraphics[width=0.97\columnwidth]{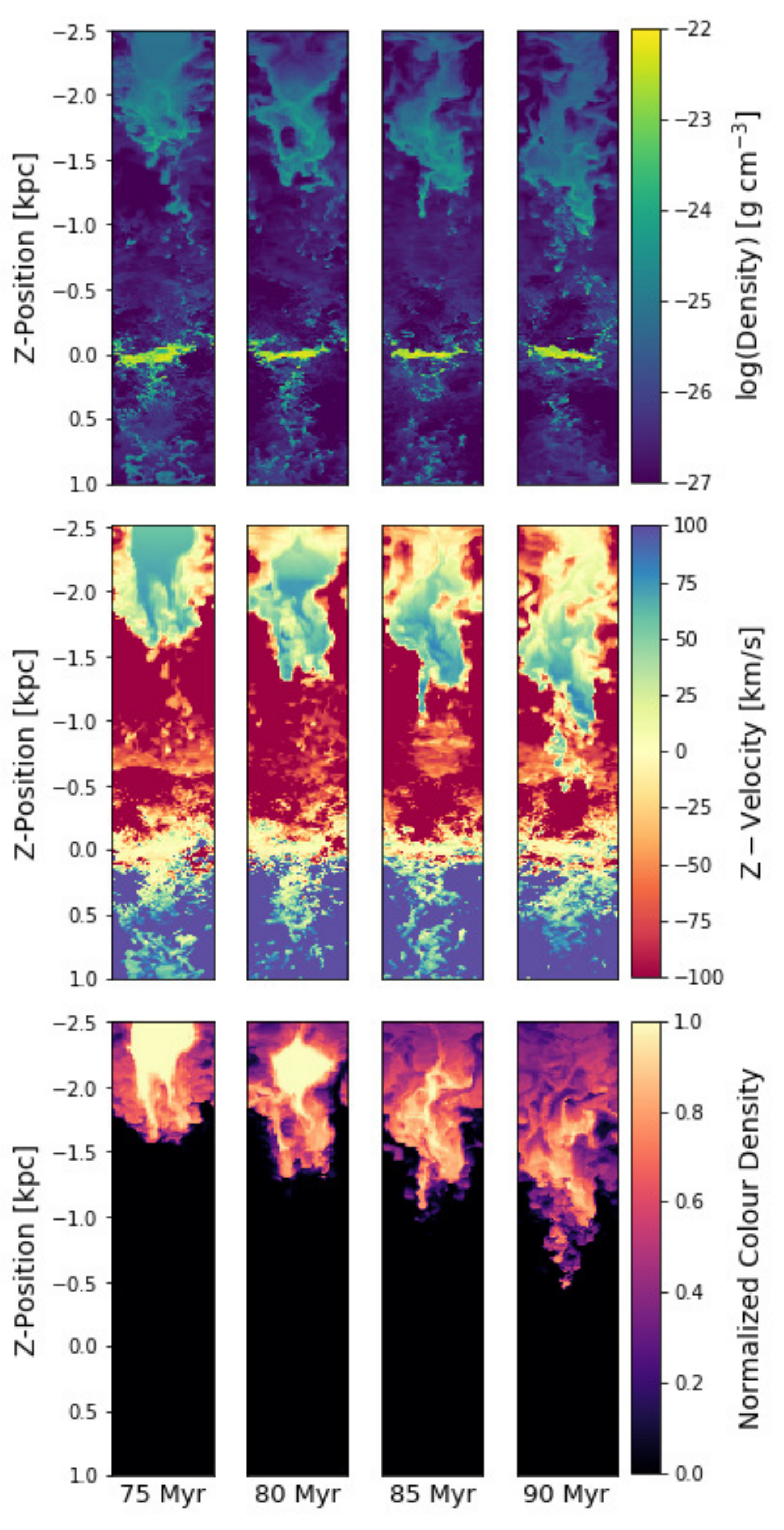}
\caption{Density, z-velocity, and normalized colour density slices showing the time evolution of the SMALL50 simulation. Each panel is a slice in the y-z plane at x = 0 with dimensions [0.35 x 3.5] kpc. Note that z-axis is vertical and each slice is only showing the top $70\%$ of the box, as the full simulation is 5 kpc in the z-direction. The time step for each panel is noted along the bottom of the plot. The plane of the disk is the yellow high-density region at z = 0. Gas is injected at the top of the box over a quarter of the surface area, and can be seen between z $\approx -2.5$ kpc to z $ \approx -1.5$ kpc at the first time step. This injected gas is given the same initial conditions as the BURST50 simulation. The normalized colour density field represents the fraction of injected gas present in a cell volume. We have changed the aspect ratio in these slices (compressing them along the z-direction) to ensure they fit in this figure.}
\label{fig:small50}
\end{figure}

In the simulations above, inflowing gas was injected across the entire cross-sectional area of the top z-boundary. This input geometry is somewhat arbitrary, and we have completed an additional simulation with a substantially smaller cloud to ensure that the extent of the cloud does not significantly influence the physical results. This small cloud simulation (SMALL50) uses the same density, temperature, and injection velocity as the BURST50 run. We inject gas from a smaller square region in the center of the top z-boundary amounting to a quarter of the total box cross-sectional area. The rest of the area on the outer edges of the top z-boundary remains set to outflow. For consistency with the other runs, we inject this cloud until it reaches a kpc in length at which time we revert the full cross-sectional area of the top z-boundary to outflow. Due to the reduced cloud extent, the resulting injected cloud mass is only about a quarter of the BURST50 injected cloud mass. The time evolution of the SMALL50 simulation in density, z-velocity, and normalized colour density is shown in Figure \ref{fig:small50}. We run this simulation to 90 Myrs, by which time the cloud is clearly on its way to interact with the disk. It has fragmented into smaller clumps and descended to within a kpc of the disk midplane.

The structure and evolution of this smaller injected cloud is similar to that of the injected gas in the BURST50 simulation. The injected gas forms a high-density front in contact with the outflowing gas and fragments into dense cold clumps. The maximum z-position of the injected gas in the SMALL50 simulation slightly lags the BURST50 injected gas at each time step. Injected gas in the BURST50 simulation was able to penetrate down towards the disk through gaps in the outflow along the edge of the simulation box where gas has not been injected in the SMALL50 run. The phase diagrams for the SMALL50 simulation have an analogous distribution to the phase diagram of the BURST50 simulation. As in the BURST50 simulation, the evolution of the SMALL50 phase diagrams show the majority of the injected gas redistributing to lower temperatures and higher densities over time. At equivalent time steps, the SMALL50 injected gas is not as close to the disk as the BURST50 simulation. Accordingly, the phase diagram of the SMALL50 injected gas has not fully populated the higher pressure isobars that BURST50 injected gas achieves at lower latitudes. However, this is largely a timing issue (the SMALL50 gas infall occurs more slowly because it was injected directly into a stream of outflowing gas at the box center) and as time goes on, it evolves in the same manner as the BURST50 run.

One notable difference in this geometry is that the outflowing gas is able to pass around the injected cloud creating shearing between the edge of the injected cloud and the adjacent outflowing gas. This shearing strips away a small fraction of the injected cloud mass. Most of this stripped material sits at high-latitudes stalled behind the injected cloud, but some ($<4\%$ of the total cloud mass) is pushed out of the box with the outflowing gas. The normalized colour density field (Figure \ref{fig:small50}) shows mixing along the shear layer, with the high-latitude stripped gas exhibiting a purity of less than $60\%$.

\subsection{Comparison to previous work}
\label{subsec:compare}
We are not the first to explore the evolution and fate of gas flowing into a galaxy. \citet{Heitsch2009} examined the evolution of similar gas inflow both in simplified wind-tunnel experiments and in hydrostatic stratified atmospheres without outflows. They found that injected gas clouds rapidly generated head-tail morphologies and ablated, generally surviving only a few kpc. This result was dependent on the size of the cloud, with larger clouds exhibiting longer survival times over greater distances. They note the importance of cooling, as some of the warm ionized material stripped from the infalling clouds was later able to condense and re-cool close to the disk. Our results are not inconsistent with this work. The closest comparison is with their larger mass, stratified cases where they find similar cloud fragmentation and cooling due the compression of these cloud fragments on their way through the halo. This cooling increases the HI mass and helps stabilize the clouds against disruption, allowing them to travel greater distances. In agreement with our cloud velocity values, they also showed an early phase of  near-ballistic acceleration followed by deceleration when they reached velocities of  $\sim 100 \rm \ km \ s^{-1}$. 
 
Similar results were also found in a set of simulations carried out using the FLASH code \citep{Kwak2011, Shelton2012, Gritton2014, Gritton2017}, where they studied individual cold clouds interacting with a hot atmosphere. Outflows in these simulations were generally implemented using the constant density and velocity wind-tunnel model. These studies also concluded that low mass clouds were quickly destroyed, with larger clouds exhibiting longer survival times. Additionally, they note the importance of cooling and mixing in models of massive clouds ($10^{5} - 10^{8} \rm M_{\odot}$), which increase their HI mass on their way to the disk by accreting a substantial amount of ambient halo gas through condensation. 

In addition, \citet{Marinacci2010} and \citet{Marinacci2011} examined the evolution of clouds moving through the hot halo near the disk under similar conditions to those examined here. However,  in these simulations the clouds were presumed to be fountain flows, the motion was not primarily vertical, and the hot halo was uniform and static. They carried out two-dimensional simulations, likewise reporting that the clouds compress, fragment, and cool efficiency thorough radiative cooling. They also concluded that cold clouds interacting with a hot halo medium will gain mass from their surroundings and decelerate due to the transfer of momentum that accompanies this exchange of material. Thermal conduction was included in the \cite{Marinacci2011} simulations by\citet{Armillotta2016}, who found that the conduction in higher temperature halos (above $4 \times 10^6$ K) suppressed hydrodynamic and thermal instabilities and reduced the importance of re-cooling.

\subsection{Simulation Caveats}
\label{subsec:caveats}
{\it Inflow conditions:}  As noted earlier, our inflow conditions are quite simplified. The constant density and velocity of material injected across the upper boundary condition of the simulation box corresponds to large, diffuse clouds. In addition, we have assumed they are relatively cold. Although these bulk conditions are motivated by cosmological simulations, their smooth injection is probably unrealistic. More likely, the clouds would have internal structure related to velocity, density and temperature perturbations (including turbulence). We argue that this would probably only increase the likelihood for the clouds to fragment into smaller clumps, and so strengthen the results that we find here. However, we recognize that very strong turbulent velocities or greatly enhanced temperatures may provide a source of non-thermal pressure and change the cloud evolution entirely.

{\it Radiative transfer:}  We make quite simplified assumptions about the radiative conditions in the cloud. In particular, we assume optically thin radiative cooling (likely reasonable given the densities) and a uniform photo-heating rate discussed in more detail in \cite{Li2017}. In reality, some ionizing radiation probably leaks out of the disk and may heat the gas \citep[e.g.,][]{BlandHawthorn1999,Dove2000}. Also, radiation pressure from both ionizing and infrared radiation could provide additional lift for the clouds \citep{Murray2010}. We suspect both of these effects will be relatively minor in determining the dynamics of the clouds in this simulation. Although heating of low temperature clouds may be important in some cases, radiation pressure is almost certainly sub-dominant to thermal and ram pressure.

{\it Magnetic fields:} There is only limited information on the magnetic field strength in clouds above the disk; however, fields are certainly observed to have strengths similar to the thermal pressure in the interstellar medium \citep[e.g.,][]{Heiles2005}. Numerical models of the disk which do include fields \citep[e.g.,][]{deAvillez2005,Hill2012,Walch2015} generally find their dynamical impact to be relatively mild, although there is some indication they may provide additional support for the vertical scale height. If important in inflowing clouds, they may help to prevent fragmentation by providing dynamical support, as well as preventing or slowing the development of the Kelvin-Helmholtz or Rayleigh-Taylor instabilities. 

{\it Numerical Resolution:} The numerical resolution in our simulations (2/4/8 pc within 0.5/1.0/2.5 kpc of the mid-plane) was chosen to be sufficiently high to resolve the Sedov phase of our SNe explosions within all but the highest density regions \citet{Li2017,Joung2012}. However, the cold clouds in the multiphase medium are not fully resolved. Indeed, without the addition of thermal conduction it is unlikely that we can fully resolve the cloud properties at any resolution. Instead they will continue to break up into smaller clumps down to the grid scale \citep{Choi2012}. 

% -----------------------------------------
\section{Summary} \label{sec:sum}

We have carried out a set of numerical simulations modeling a cloud of gas inflowing towards the galactic disk from a height of 2.5 kpc. We include realistic heating and outflows in order to better understand gas interactions at the disk-halo interface. Our galactic model is taken from \citet{Li2017} and is aimed at approximating conditions similar to the local neighborhood of the Milky Way. We include SN explosions resolved with a 2 pc spatial grid, which should be sufficient to model their energetic input. The gas accretion is simulated with a smooth inflow of gas at the upper boundary condition. We explore three models which vary in the velocity of the inflow, ranging from 10 to 100 $\rm km \ s^{-1}$. In each case, a similar amount of total gas is injected before the inflow is curtailed and the resulting flow is evolved for ~50-100 Myr. We summarize the following key results:

\begin{enumerate}

\item{\it Cloud Evolution} The initial diffuse cloud interacts with the outflowing gas, forming a forward shock, contact discontinuity and reverse shock. Most of the cloud mass ends up behind the discontinuity, cooling and fragmenting via Rayleigh Taylor and Kelvin Helmholtz instabilities. The clumps have typical densities of 1 cm$^{-3}$ and temperatures of about 1000 K. The dense clumps can accrete onto the disk even when the ram pressure of the outflow exceeds that of the original diffuse cloud (Figures~\ref{fig:initial}, \ref{fig:panels} and \ref{fig:phase}).

\item{\it Cloud Velocity and Mixing}  The clouds initially accelerate ballistically before being decelerated by interactions with the outflowing material. Close to the disk, we find typical cloud velocities ranging from 40-100 $\rm km \ s^{-1}$, depending on their initial injection rate. This approximately corresponds to intermediate velocity clouds. The deceleration appears to be largely controlled by mixing, with velocities of dense gas correlating strongly with the amount of mixing the cloud has experienced. Clouds do mix somewhat with the ambient medium but retain more than 50\% of their original gas (Figures~\ref{fig:profiles} and \ref{fig:phase}).

\item{\it Cloud Fate and Accretion} In these models, infalling gas clouds often penetrate through the disk, either because they hit an area with low disk surface density, or because they have a high ram pressure. Most of the penetrating gas is reaccelerated by the outflow on the other side of the disk and expelled. Only in the case with the lowest initial velocity (10 km/s) does a majority of the cloud mass end up accreting on to the disk (Figure~\ref{fig:fate}).

\item{\it Observational comparison} We generate mock observables for HI (a low ionization ion) and OVI (a high ionization ion). We predict that inflow produces significant HI column densities of $\sim 10^{20}$ cm$^{-2}$ in the compressed injected gas, with even higher values for simulations with higher inflow rates. This is consistent with observational column densities of IVC's as discussed in section \ref{sec:density}. We find correspondingly high H$\alpha$ emission in the high-density (reverse shock) region with an elevated emission intensity of  $[1-15] \times 10^{4} \rm \ ph \ s^{-1} \ cm^{-2} \ sr^{-1}$. Closer to the disk, emission from the fragmented cloud structures is difficult to distinguish from the background outflow emission intensity. Because the clouds stay largely cool, we find that there is relatively little OVI absorption or emission. We make comparisons to observations, finding rough agreement with many existing Milky Way measurements. (Figures~\ref{fig:col} and \ref{fig:emis}).
\end{enumerate}

This work highlights the importance of interactions between inflowing gas clouds and outflowing gas at the disk-halo interface. The compression, fragmentation and cooling of inflowing gas results in the formation of cold, dense clouds which descend to the disk (and beyond) without being destroyed. Future work is required to explore a wider parameter space, both of cloud properties and of disk conditions. In addition, we have neglected a number of physical effects, the importance of which should be examined in later work.

\acknowledgements
We thank Mary Putman, David Schiminovich, Yong Zheng and our referee for their useful discussions and thoughtful feedback, which have helped us improve various drafts of this manuscript. Many thanks to Lauren Corlies for sharing her {\sc Cloudy} models used in our mock observations. We acknowledge support from NSF grants AST-1312888, AST-1615955 and DGE-1644869, and NASA grant NNX15AB20G, as well as computational resources from NSF XSEDE, and Columbia University's compute cluster. Simulations were run on the Texas Advanced Computing Center Stampede supercomputer and were performed using the publicly-available Enzo code (http://enzo-project.org) and analyzed with the yt package \citep{Turk2011}. Enzo is the product of a collaborative effort of many independent scientists from numerous institutions around the world. Their commitment to open science has helped make this work possible. The Flatiron Institute is supported by the generosity of the Simons Foundation.

\section*{Appendix}
Movies for the BURST50, BURST100, and IMD10 simulations can be found at the following url: \url{https://doi.org/10.7916/d8-vgnn-7n73}. They are also embedded in the Appendix of the ApJ publication of this work. We have included the corresponding still frames (the first frame of each movie) below. In previous figures we changed the aspect ratio of the simulation slices (compressing them along the z-direction) to ensure they fit in the figure. In these movies/slices we maintain the correct aspect ratio (10:1). 

%\clearpage

\begin{figure}
{\includegraphics[width = \columnwidth]{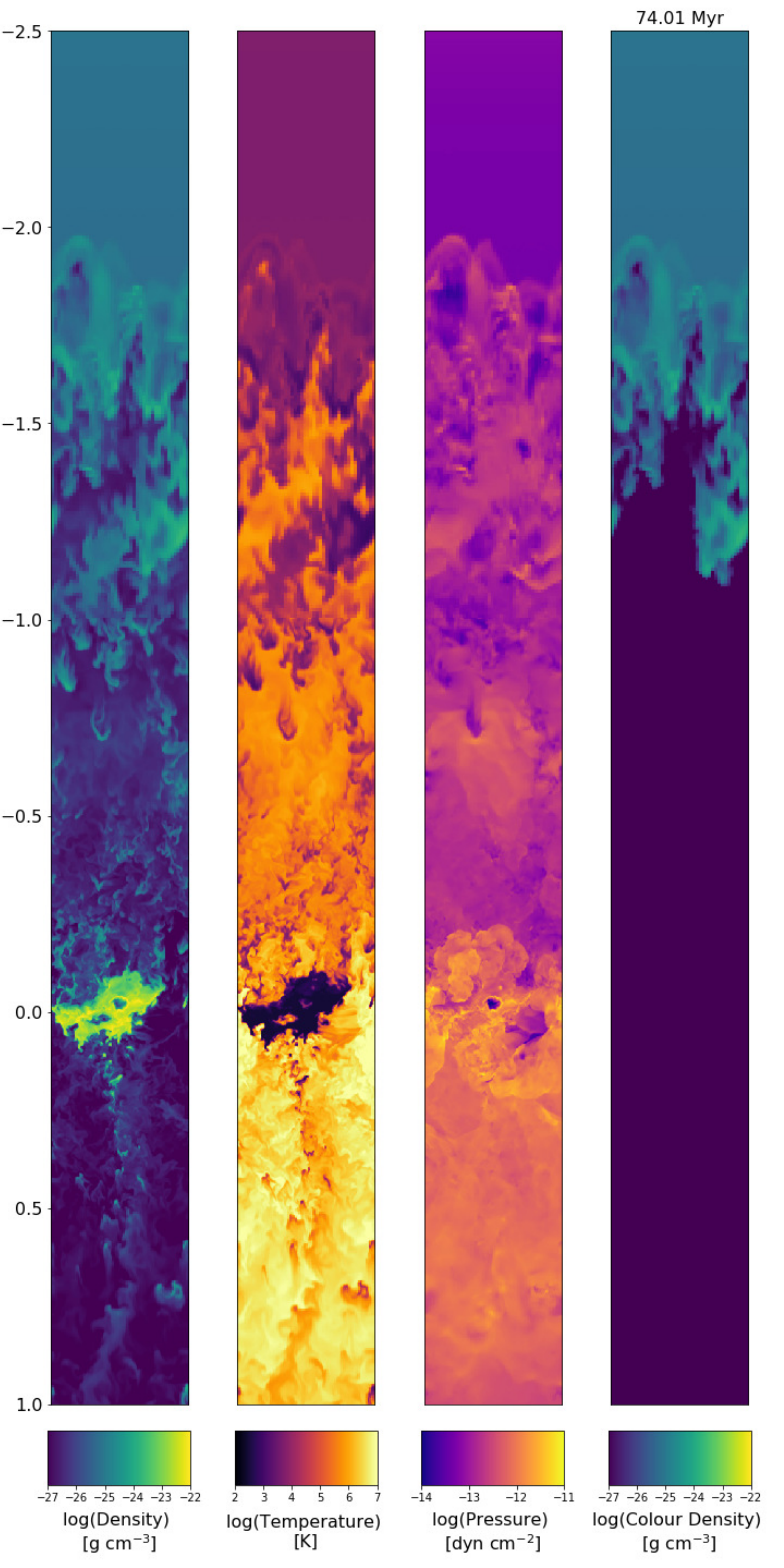}}
\caption{A movie of the BURST50 simulation showing the evolution of density, temperature, pressure, and colour density as a function of time. Each panel is a slice in the y-z plane at x = 0 with dimensions [0.35 x 3.5] kpc. Note that z-axis is vertical and each slice is only showing the top $70\%$ of the box, as the full simulation is 5 kpc in the z-direction. The time step is noted in the top right corner. The plane of the disk is the yellow high-density region at z = 0. Gas is injected at the top of the box, and can be seen independent from the surrounding gas using the colour density field. The colour density traces the density of only the injected gas, while the rest of the gas in the domain starts with a near-zero colour density field. This injected gas is given an initial density of $8.35 \times 10^{-26} \rm g \ cm^{-3}$, an initial temperature of $\sim 10^{4}$ K, and an initial velocity of 50 km $\rm s^{-1}$. We have maintained the correct aspect ratio in these slices.}
\label{fig:still50}
\end{figure}

\begin{figure}
{\includegraphics[width = \columnwidth]{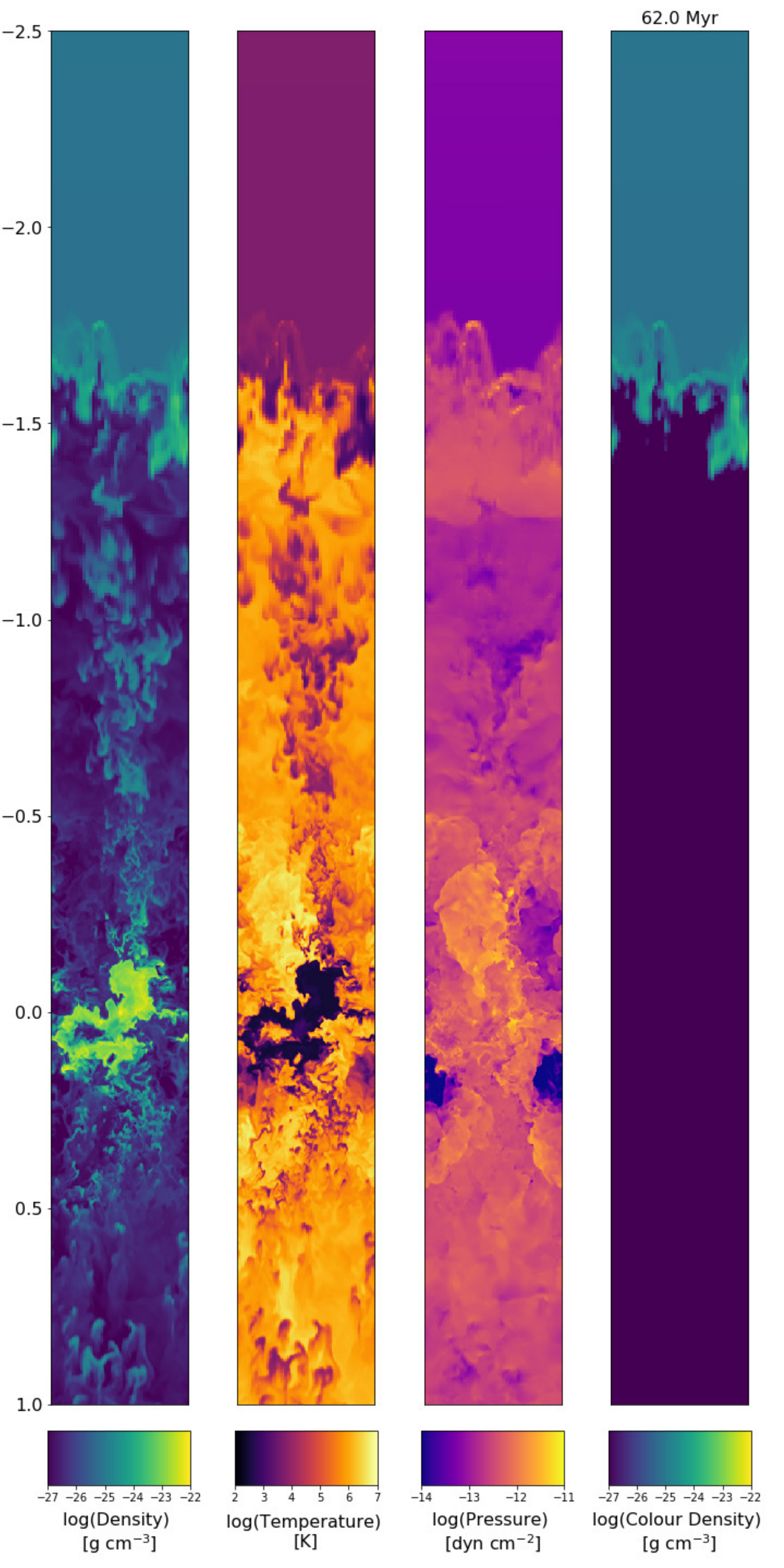}}
\caption{A movie of the BURST100 simulation showing the evolution of density, temperature, pressure, and colour density as a function of time. Each panel is a slice in the y-z plane at x = 0 with dimensions [0.35 x 3.5] kpc. Note that z-axis is vertical and each slice is only showing the top $70\%$ of the box, as the full simulation is 5 kpc in the z-direction. The time step is noted in the top right corner. The plane of the disk is the yellow high-density region at z = 0. Gas is injected at the top of the box, and can be seen independent from the surrounding gas using the colour density field. The colour density traces the density of only the injected gas, while the rest of the gas in the domain starts with a near-zero colour density field. This injected gas is given an initial density of $8.35 \times 10^{-26} \rm g \ cm^{-3}$, an initial temperature of $\sim 10^{4}$ K, and an initial velocity of 100 km $\rm s^{-1}$. We have maintained the correct aspect ratio in these slices.} 
\label{fig:still100}
\end{figure}

\begin{figure}
{\includegraphics[width = \columnwidth]{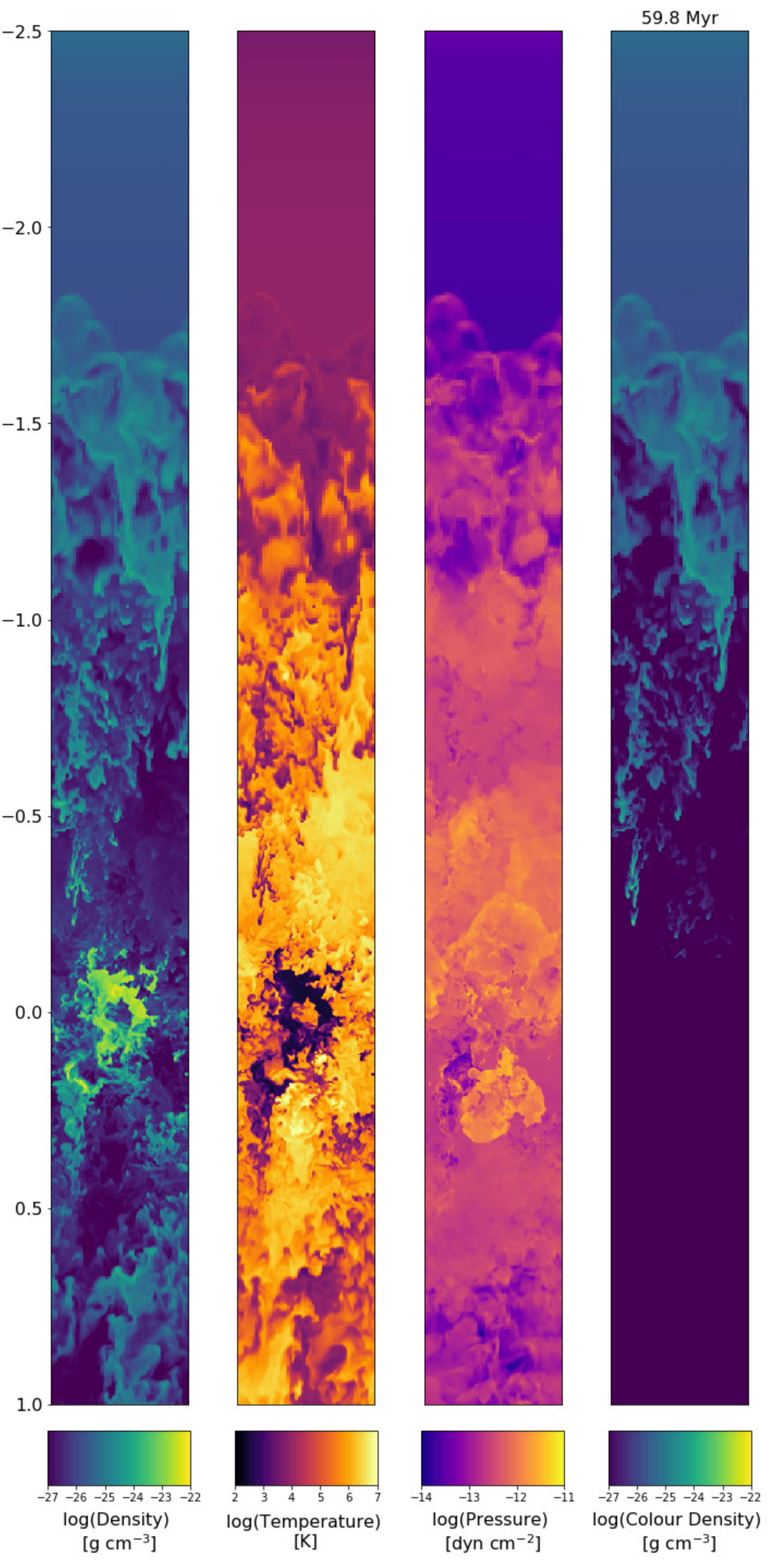}}
\caption{A movie of the IMD10 simulation showing the evolution of density, temperature, pressure, and colour density as a function of time. Each panel is a slice in the y-z plane at x = 0 with dimensions [0.35 x 3.5] kpc. Note that z-axis is vertical and each slice is only showing the top $70\%$ of the box, as the full simulation is 5 kpc in the z-direction. The time step is noted in the top right corner. The plane of the disk is the yellow high-density region at z = 0. Gas is injected at the top of the box, and can be The colour density traces the density of only the injected gas, while the rest of the gas in the domain starts with a near-zero colour density field. The colour density traces the density of only the injected gas, while the rest of the gas in the domain starts with a near-zero colour density field. This injected gas is given an initial density of $8.35 \times 10^{-26} \rm g \ cm^{-3}$, an initial temperature of $\sim 10^{4}$ K, and an initial velocity of 10 km $\rm s^{-1}$. We have maintained the correct aspect ratio in these slices.} 
\label{fig:still10}
\end{figure}


\newpage
\bibliography{mybib}

\begin{thebibliography}{}
\expandafter\ifx\csname natexlab\endcsname\relax\def\natexlab#1{#1}\fi
\providecommand{\url}[1]{\href{#1}{#1}}

\bibitem[{{Angl{\'e}s-Alc{\'a}zar} {et~al.}(2017){Angl{\'e}s-Alc{\'a}zar},
  {Faucher-Gigu{\`e}re}, {Kere{\v s}}, {Hopkins}, {Quataert}, \&
  {Murray}}]{AA2017}
{Angl{\'e}s-Alc{\'a}zar}, D., {Faucher-Gigu{\`e}re}, C.-A., {Kere{\v s}}, D.,
  {et~al.} 2017, \mnras, 470, 4698

\bibitem[{{Armillotta} {et~al.}(2016){Armillotta}, {Fraternali}, \&
  {Marinacci}}]{Armillotta2016}
{Armillotta}, L., {Fraternali}, F., \& {Marinacci}, F. 2016, \mnras, 462, 4157

\bibitem[{{Barstow} {et~al.}(2010){Barstow}, {Boyce}, {Welsh}, {Lallement},
  {Barstow}, {Forbes}, \& {Preval}}]{Barstow2010}
{Barstow}, M.~A., {Boyce}, D.~D., {Welsh}, B.~Y., {et~al.} 2010, \apj, 723,
  1762

\bibitem[{{Bizyaev} {et~al.}(2017){Bizyaev}, {Walterbos}, {Yoachim}, {Riffel},
  {Fern{\'a}ndez-Trincado}, {Pan}, {Diamond-Stanic}, {Jones}, {Thomas},
  {Cleary}, \& {Brinkmann}}]{Bizyaev2017}
{Bizyaev}, D., {Walterbos}, R.~A.~M., {Yoachim}, P., {et~al.} 2017, \apj, 839,
  87

\bibitem[{{Bland-Hawthorn} \& {Maloney}(1999)}]{BlandHawthorn1999}
{Bland-Hawthorn}, J., \& {Maloney}, P.~R. 1999, \apjl, 510, L33

\bibitem[{{Bowen} {et~al.}(2008){Bowen}, {Jenkins}, {Tripp}, {Sembach},
  {Savage}, {Moos}, {Oegerle}, {Friedman}, {Gry}, {Kruk}, {Murphy}, {Sankrit},
  {Shull}, {Sonneborn}, \& {York}}]{Bowen2008}
{Bowen}, D.~V., {Jenkins}, E.~B., {Tripp}, T.~M., {et~al.} 2008, The
  Astrophysical Journal Supplement Series, 176, 59

\bibitem[{{Bregman}(1980)}]{Bregman1980}
{Bregman}, J.~N. 1980, \apj, 236, 577

\bibitem[{{Bryan} {et~al.}(2014){Bryan}, {Norman}, {O'Shea}, {Abel}, {Wise},
  {Turk}, {Reynolds}, {Collins}, {Wang}, {Skillman}, {Smith}, {Harkness},
  {Bordner}, {Kim}, {Kuhlen}, {Xu}, {Goldbaum}, {Hummels}, {Kritsuk}, {Tasker},
  {Skory}, {Simpson}, {Hahn}, {Oishi}, {So}, {Zhao}, {Cen}, {Li}, \& {Enzo
  Collaboration}}]{2014ApJS..211...19B}
{Bryan}, G.~L., {Norman}, M.~L., {O'Shea}, B.~W., {et~al.} 2014, The
  Astrophysical Journal Supplement Series, 211, doi:10.1088/0067-0049/211/2/19

\bibitem[{{Cen} \& {Ostriker}(1999)}]{1999ApJ...514....1C}
{Cen}, R., \& {Ostriker}, J.~P. 1999, \apj, 514, 1

\bibitem[{{Chiappini} {et~al.}(2001){Chiappini}, {Matteucci}, \&
  {Romano}}]{Chiappini2001}
{Chiappini}, C., {Matteucci}, F., \& {Romano}, D. 2001, \apj, 554, 1044

\bibitem[{{Choi} \& {Stone}(2012)}]{Choi2012}
{Choi}, E., \& {Stone}, J.~M. 2012, \apj, 747, 86

\bibitem[{{Chomiuk} \& {Povich}(2011)}]{Chomiuk2011}
{Chomiuk}, L., \& {Povich}, M.~S. 2011, \aj, 142, 197

\bibitem[{{Corlies} \& {Schiminovich}(2016)}]{2016ApJ...827..148C}
{Corlies}, L., \& {Schiminovich}, D. 2016, \apj, 827,
  doi:10.3847/0004-637X/827/2/148

\bibitem[{{Creasey} {et~al.}(2013){Creasey}, {Theuns}, \&
  {Bower}}]{Creasey2013}
{Creasey}, P., {Theuns}, T., \& {Bower}, R.~G. 2013, \mnras, 429, 1922

\bibitem[{{de Avillez} \& {Breitschwerdt}(2005)}]{deAvillez2005}
{de Avillez}, M.~A., \& {Breitschwerdt}, D. 2005, \aap, 436, 585

\bibitem[{{Dickey} \& {Lockman}(1990)}]{Dickey1990}
{Dickey}, J.~M., \& {Lockman}, F.~J. 1990, Annual Review of Astronomy and
  Astrophysics, 28, 215

\bibitem[{{Dove} {et~al.}(2000){Dove}, {Shull}, \& {Ferrara}}]{Dove2000}
{Dove}, J.~B., {Shull}, J.~M., \& {Ferrara}, A. 2000, \apj, 531, 846

\bibitem[{{Erb}(2008)}]{Erb2008}
{Erb}, D.~K. 2008, \apj, 674, 151

\bibitem[{{Fenner} \& {Gibson}(2003)}]{Fenner2003}
{Fenner}, Y., \& {Gibson}, B.~K. 2003, Publications of the Astronomical Society
  of Australia, 20, 189

\bibitem[{{Ferland} {et~al.}(1998){Ferland}, {Korista}, {Verner}, {Ferguson},
  {Kingdon}, \& {Verner}}]{1998PASP..110..761F}
{Ferland}, G.~J., {Korista}, K.~T., {Verner}, D.~A., {et~al.} 1998,
  Publications of the Astronomical Society of the Pacific, 110, 761

\bibitem[{{Fern{\'a}ndez} {et~al.}(2012){Fern{\'a}ndez}, {Joung}, \&
  {Putman}}]{Fernandez2012}
{Fern{\'a}ndez}, X., {Joung}, M.~R., \& {Putman}, M.~E. 2012, \apj, 749, 181

\bibitem[{{Fielding} {et~al.}(2018){Fielding}, {Quataert}, \&
  {Martizzi}}]{Fielding2018}
{Fielding}, D., {Quataert}, E., \& {Martizzi}, D. 2018, \mnras, 481, 3325

\bibitem[{{Ford} {et~al.}(2010){Ford}, {Lockman}, \&
  {McClure-Griffiths}}]{Ford2010}
{Ford}, H.~A., {Lockman}, F.~J., \& {McClure-Griffiths}, N.~M. 2010, \apj, 722,
  367

\bibitem[{{Fraternali} \& {Binney}(2008)}]{Fraternali2008}
{Fraternali}, F., \& {Binney}, J.~J. 2008, \mnras, 386, 935

\bibitem[{{Fraternali} {et~al.}(2002){Fraternali}, {van Moorsel}, {Sancisi}, \&
  {Oosterloo}}]{Fraternali2002}
{Fraternali}, F., {van Moorsel}, G., {Sancisi}, R., \& {Oosterloo}, T. 2002,
  \aj, 123, 3124

\bibitem[{{Gaensler} {et~al.}(2008){Gaensler}, {Madsen}, {Chatterjee}, \&
  {Mao}}]{Gaensler2008}
{Gaensler}, B.~M., {Madsen}, G.~J., {Chatterjee}, S., \& {Mao}, S.~A. 2008,
  Publications of the Astronomical Society of Australia, 25, 184

\bibitem[{{Gritton} {et~al.}(2017){Gritton}, {Shelton}, \&
  {Galyardt}}]{Gritton2017}
{Gritton}, J.~A., {Shelton}, R.~L., \& {Galyardt}, J.~E. 2017, \apj, 842, 102

\bibitem[{{Gritton} {et~al.}(2014){Gritton}, {Shelton}, \&
  {Kwak}}]{Gritton2014}
{Gritton}, J.~A., {Shelton}, R.~L., \& {Kwak}, K. 2014, \apj, 795, 99

\bibitem[{{Haffner} {et~al.}(2003){Haffner}, {Reynolds}, {Tufte}, {Madsen},
  {Jaehnig}, \& {Percival}}]{Haffner2003}
{Haffner}, L.~M., {Reynolds}, R.~J., {Tufte}, S.~L., {et~al.} 2003, The
  Astrophysical Journal Supplement Series, 149, 405

\bibitem[{{Heald} {et~al.}(2011){Heald}, {J{\'o}zsa}, {Serra}, {Zschaechner},
  {Rand}, {Fraternali}, {Oosterloo}, {Walterbos}, {J{\"u}tte}, \&
  {Gentile}}]{Heald2011}
{Heald}, G., {J{\'o}zsa}, G., {Serra}, P., {et~al.} 2011, \aap, 526, A118

\bibitem[{{Heald} {et~al.}(2006){Heald}, {Rand}, {Benjamin}, \&
  {Bershady}}]{Heald2006}
{Heald}, G.~H., {Rand}, R.~J., {Benjamin}, R.~A., \& {Bershady}, M.~A. 2006,
  \apj, 647, 1018

\bibitem[{{Heald} {et~al.}(2007){Heald}, {Rand}, {Benjamin}, \&
  {Bershady}}]{Heald2007}
---. 2007, \apj, 663, 933

\bibitem[{{Heiles} \& {Crutcher}(2005)}]{Heiles2005}
{Heiles}, C., \& {Crutcher}, R. 2005, in Lecture Notes in Physics, Berlin
  Springer Verlag, Vol. 664, Cosmic Magnetic Fields, ed. R.~{Wielebinski} \&
  R.~{Beck}, 137

\bibitem[{{Heitsch} \& {Putman}(2009)}]{Heitsch2009}
{Heitsch}, F., \& {Putman}, M.~E. 2009, \apj, 698, 1485

\bibitem[{{Hill} {et~al.}(2009){Hill}, {Haffner}, \& {Reynolds}}]{Hill2009}
{Hill}, A.~S., {Haffner}, L.~M., \& {Reynolds}, R.~J. 2009, \apj, 703, 1832

\bibitem[{{Hill} {et~al.}(2012){Hill}, {Joung}, {Mac Low}, {Benjamin},
  {Haffner}, {Klingenberg}, \& {Waagan}}]{Hill2012}
{Hill}, A.~S., {Joung}, M.~R., {Mac Low}, M.-M., {et~al.} 2012, \apj, 750, 104

\bibitem[{{Hill} {et~al.}(2013){Hill}, {Mao}, {Benjamin}, {Lockman}, \&
  {McClure-Griffiths}}]{Hill2013}
{Hill}, A.~S., {Mao}, S.~A., {Benjamin}, R.~A., {Lockman}, F.~J., \&
  {McClure-Griffiths}, N.~M. 2013, \apj, 777, 55

\bibitem[{{Hobbs} {et~al.}(2013){Hobbs}, {Read}, {Power}, \&
  {Cole}}]{Hobbs2013}
{Hobbs}, A., {Read}, J., {Power}, C., \& {Cole}, D. 2013, \mnras, 434, 1849

\bibitem[{{Hopkins} {et~al.}(2008){Hopkins}, {McClure-Griffiths}, \&
  {Gaensler}}]{Hopkins2008}
{Hopkins}, A.~M., {McClure-Griffiths}, N.~M., \& {Gaensler}, B.~M. 2008, \apj,
  682, L13

\bibitem[{{Houck} \& {Bregman}(1990)}]{Houck1990}
{Houck}, J.~C., \& {Bregman}, J.~N. 1990, \apj, 352, 506

\bibitem[{{Jenkins}(1978)}]{Jenkins1978b}
{Jenkins}, E.~B. 1978, \apj, 219, 845

\bibitem[{{Joung} \& {Mac Low}(2006)}]{Joung2006}
{Joung}, M.~K.~R., \& {Mac Low}, M.-M. 2006, \apj, 653, 1266

\bibitem[{{Joung} {et~al.}(2012{\natexlab{a}}){Joung}, {Bryan}, \&
  {Putman}}]{Joung2012b}
{Joung}, M.~R., {Bryan}, G.~L., \& {Putman}, M.~E. 2012{\natexlab{a}}, \apj,
  745, 148

\bibitem[{{Joung} {et~al.}(2012{\natexlab{b}}){Joung}, {Putman}, {Bryan},
  {Fern{\'a}ndez}, \& {Peek}}]{Joung2012}
{Joung}, M.~R., {Putman}, M.~E., {Bryan}, G.~L., {Fern{\'a}ndez}, X., \&
  {Peek}, J.~E.~G. 2012{\natexlab{b}}, \apj, 759, 137

\bibitem[{{Kalberla} {et~al.}(2005){Kalberla}, {Burton}, {Hartmann}, {Arnal},
  {Bajaja}, {Morras}, \& {P{\"o}ppel}}]{2005A&A...440..775K}
{Kalberla}, P.~M.~W., {Burton}, W.~B., {Hartmann}, D., {et~al.} 2005, \aap,
  440, 775

\bibitem[{{Kennicutt}(1989)}]{1989ApJ...344..685K}
{Kennicutt}, Robert~C., J. 1989, \apj, 344, 685

\bibitem[{{Kere{\v s}} {et~al.}(2009){Kere{\v s}}, {Katz}, {Fardal},
  {Dav{\'e}}, \& {Weinberg}}]{Keres2009}
{Kere{\v s}}, D., {Katz}, N., {Fardal}, M., {Dav{\'e}}, R., \& {Weinberg},
  D.~H. 2009, \mnras, 395, 160

\bibitem[{{Kere{\v{s}}} {et~al.}(2005){Kere{\v{s}}}, {Katz}, {Weinberg}, \&
  {Dav{\'e}}}]{Keres2005}
{Kere{\v{s}}}, D., {Katz}, N., {Weinberg}, D.~H., \& {Dav{\'e}}, R. 2005,
  \mnras, 363, 2

\bibitem[{{Kim} {et~al.}(2017){Kim}, {Ostriker}, \& {Raileanu}}]{Kim2017}
{Kim}, C.-G., {Ostriker}, E.~C., \& {Raileanu}, R. 2017, \apj, 834, 25

\bibitem[{{Krishnarao} {et~al.}(2017){Krishnarao}, {Haffner}, {Benjamin},
  {Hill}, \& {Barger}}]{2017ApJ...838...43K}
{Krishnarao}, D., {Haffner}, L.~M., {Benjamin}, R.~A., {Hill}, A.~S., \&
  {Barger}, K.~A. 2017, \apj, 838, 43

\bibitem[{{Kwak} {et~al.}(2011){Kwak}, {Henley}, \& {Shelton}}]{Kwak2011}
{Kwak}, K., {Henley}, D.~B., \& {Shelton}, R.~L. 2011, \apj, 739, 30

\bibitem[{{Leitner} \& {Kravtsov}(2011)}]{Leitner2011}
{Leitner}, S.~N., \& {Kravtsov}, A.~V. 2011, \apj, 734, 48

\bibitem[{{Leroy} {et~al.}(2013){Leroy}, {Walter}, {Sandstrom}, {Schruba},
  {Munoz-Mateos}, {Bigiel}, {Bolatto}, {Brinks}, {de Blok}, {Meidt}, {Rix},
  {Rosolowsky}, {Schinnerer}, {Schuster}, \& {Usero}}]{Leroy2013}
{Leroy}, A.~K., {Walter}, F., {Sandstrom}, K., {et~al.} 2013, \aj, 146, 19

\bibitem[{{Levine} {et~al.}(2006){Levine}, {Blitz}, \& {Heiles}}]{Levine2006}
{Levine}, E.~S., {Blitz}, L., \& {Heiles}, C. 2006, \apj, 643, 881

\bibitem[{{Li} {et~al.}(2017{\natexlab{a}}){Li}, {Bryan}, \&
  {Ostriker}}]{Li2017}
{Li}, M., {Bryan}, G.~L., \& {Ostriker}, J.~P. 2017{\natexlab{a}}, \apj, 841,
  doi:10.3847/1538-4357/aa7263

\bibitem[{{Li} {et~al.}(2017{\natexlab{b}}){Li}, {Bryan}, \&
  {Ostriker}}]{Li2017b}
---. 2017{\natexlab{b}}, \apj, 835, L10

\bibitem[{{Lockman}(2002)}]{Lockman2002}
{Lockman}, F.~J. 2002, \apj, 580, L47

\bibitem[{{Lockman}(2015)}]{2015IAUGA..2257086L}
{Lockman}, F.~J. 2015, in IAU General Assembly, Meeting \#29, id.2257086,
  Vol.~22, 2257086

\bibitem[{{Lockman}(2016)}]{Lockman2016}
{Lockman}, F.~J. 2016, in From Interstellar Clouds to Star-Forming Galaxies:
  Universal Processes?, Proceedings of the International Astronomical Union,
  IAU Symposium, Volume 315, pp. 9-12, Vol. 315, 9--12

\bibitem[{{Lockman} {et~al.}(2008){Lockman}, {Benjamin}, {Heroux}, \&
  {Langston}}]{Lockman2008}
{Lockman}, F.~J., {Benjamin}, R.~A., {Heroux}, A.~J., \& {Langston}, G.~I.
  2008, \apj, 679, L21

\bibitem[{{Marasco} {et~al.}(2012){Marasco}, {Fraternali}, \&
  {Binney}}]{Marasco2012}
{Marasco}, A., {Fraternali}, F., \& {Binney}, J.~J. 2012, \mnras, 419, 1107

\bibitem[{{Marinacci} {et~al.}(2010){Marinacci}, {Binney}, {Fraternali},
  {Nipoti}, {Ciotti}, \& {Londrillo}}]{Marinacci2010}
{Marinacci}, F., {Binney}, J., {Fraternali}, F., {et~al.} 2010, \mnras, 404,
  1464

\bibitem[{{Marinacci} {et~al.}(2011){Marinacci}, {Fraternali}, {Nipoti},
  {Binney}, {Ciotti}, \& {Londrillo}}]{Marinacci2011}
{Marinacci}, F., {Fraternali}, F., {Nipoti}, C., {et~al.} 2011, \mnras, 415,
  1534

\bibitem[{{Miller} \& {Bregman}(2013)}]{Miller2013}
{Miller}, M.~J., \& {Bregman}, J.~N. 2013, \apj, 770, 118

\bibitem[{{Miller} \& {Bregman}(2015)}]{Miller2015}
---. 2015, \apj, 800, 14

\bibitem[{{Muratov} {et~al.}(2015){Muratov}, {Kere{\v s}},
  {Faucher-Gigu{\`e}re}, {Hopkins}, {Quataert}, \& {Murray}}]{Muratov2015}
{Muratov}, A.~L., {Kere{\v s}}, D., {Faucher-Gigu{\`e}re}, C.-A., {et~al.}
  2015, \mnras, 454, 2691

\bibitem[{{Murray} {et~al.}(2010){Murray}, {Quataert}, \&
  {Thompson}}]{Murray2010}
{Murray}, N., {Quataert}, E., \& {Thompson}, T.~A. 2010, \apj, 709, 191

\bibitem[{{Nelson} {et~al.}(2013){Nelson}, {Vogelsberger}, {Genel}, {Sijacki},
  {Kere{\v s}}, {Springel}, \& {Hernquist}}]{Nelson2013}
{Nelson}, D., {Vogelsberger}, M., {Genel}, S., {et~al.} 2013, \mnras, 429, 3353

\bibitem[{{Norman} \& {Ikeuchi}(1989)}]{Norman1989}
{Norman}, C.~A., \& {Ikeuchi}, S. 1989, \apj, 345, 372

\bibitem[{{Oegerle} {et~al.}(2005){Oegerle}, {Jenkins}, {Shelton}, {Bowen}, \&
  {Chayer}}]{Oegerle2005}
{Oegerle}, W.~R., {Jenkins}, E.~B., {Shelton}, R.~L., {Bowen}, D.~V., \&
  {Chayer}, P. 2005, \apj, 622, 377

\bibitem[{{Oosterloo} {et~al.}(2007){Oosterloo}, {Fraternali}, \&
  {Sancisi}}]{Oosterloo2007}
{Oosterloo}, T., {Fraternali}, F., \& {Sancisi}, R. 2007, \aj, 134, 1019

\bibitem[{{Oppenheimer} {et~al.}(2010){Oppenheimer}, {Dav{\'e}}, {Kere{\v s}},
  {Fardal}, {Katz}, {Kollmeier}, \& {Weinberg}}]{Oppenheimer2010}
{Oppenheimer}, B.~D., {Dav{\'e}}, R., {Kere{\v s}}, D., {et~al.} 2010, \mnras,
  406, 2325

\bibitem[{{Osterbrock}(1989)}]{Osterbrock1989}
{Osterbrock}, D.~E. 1989, {Astrophysics of gaseous nebulae and active galactic
  nuclei}

\bibitem[{{Otte} \& {Dixon}(2006)}]{Otte2006}
{Otte}, B., \& {Dixon}, W. V.~D. 2006, \apj, 647, 312

\bibitem[{{Peek} {et~al.}(2011){Peek}, {Heiles}, {Douglas}, {Lee}, {Grcevich},
  {Stanimirovi{\'c}}, {Putman}, {Korpela}, {Gibson}, {Begum}, {Saul},
  {Robishaw}, \& {Kr{\v{c}}o}}]{Peek2011}
{Peek}, J.~E.~G., {Heiles}, C., {Douglas}, K.~A., {et~al.} 2011, The
  Astrophysical Journal Supplement Series, 194, 20

\bibitem[{{Prochaska} {et~al.}(2017){Prochaska}, {Werk}, {Worseck}, {Tripp},
  {Tumlinson}, {Burchett}, {Fox}, {Fumagalli}, {Lehner}, {Peeples}, \&
  {Tejos}}]{2017ApJ...837..169P}
{Prochaska}, J.~X., {Werk}, J.~K., {Worseck}, G., {et~al.} 2017, \apj, 837, 169

\bibitem[{{Putman}(2017)}]{Putman2017}
{Putman}, M.~E. 2017, in Gas Accretion onto Galaxies, Vol. 430, 1

\bibitem[{{Putman} {et~al.}(2012){Putman}, {Peek}, \& {Joung}}]{Putman2012}
{Putman}, M.~E., {Peek}, J.~E.~G., \& {Joung}, M.~R. 2012, Annual Review of
  Astronomy and Astrophysics, 50, 491

\bibitem[{{Putman} {et~al.}(2009){Putman}, {Henning}, {Bolatto}, {Keres},
  {Pisano}, {Rosenberg}, {Bigiel}, {Bryan}, {Calzetti}, {Carilli}, {Charlton},
  {Chen}, {Darling}, {Gibson}, {Gnedin}, {Gnedin}, {Heitsch}, {Hunter},
  {Kannappan}, {Krumholz}, {Lazarian}, {Lasio}, {Leroy}, {Lockman}, {Mac Low},
  {Maller}, {Meurer}, {O'Neil}, {Ostriker}, {Peek}, {Prochaska}, {Rand},
  {Robertson}, {Schiminovich}, {Simon}, {Stanimirovic}, {Thilker}, {Thom},
  {Tinker}, {Wakker}, {Weiner}, {van der Hulst}, {Wolfe}, {Wong}, \&
  {Young}}]{Putman2009}
{Putman}, M.~E., {Henning}, P., {Bolatto}, A., {et~al.} 2009, in astro2010: The
  Astronomy and Astrophysics Decadal Survey, Vol. 2010, 241

\bibitem[{{Rand}(2000)}]{Rand2000}
{Rand}, R.~J. 2000, \apj, 537, L13

\bibitem[{{Reynolds}(1993)}]{Reynolds1993}
{Reynolds}, R.~J. 1993, in Back to the Galaxy, Vol. 278, 156--165

\bibitem[{{Rosen} \& {Bregman}(1995)}]{Rosen1995}
{Rosen}, A., \& {Bregman}, J.~N. 1995, \apj, 440, 634

\bibitem[{{Sancisi} {et~al.}(2001){Sancisi}, {Fraternali}, {Oosterloo}, \& {van
  Moorsel}}]{Sancisi2001}
{Sancisi}, R., {Fraternali}, F., {Oosterloo}, T., \& {van Moorsel}, G. 2001, in
  Gas and Galaxy Evolution, Vol. 240, 241

\bibitem[{{Saul} {et~al.}(2012){Saul}, {Peek}, {Grcevich}, {Putman}, {Douglas},
  {Korpela}, {Stanimirovi{\'c}}, {Heiles}, {Gibson}, {Lee}, {Begum}, {Brown},
  {Burkhart}, {Hamden}, {Pingel}, \& {Tonnesen}}]{Saul2012}
{Saul}, D.~R., {Peek}, J.~E.~G., {Grcevich}, J., {et~al.} 2012, \apj, 758, 44

\bibitem[{{Savage} \& {Lehner}(2006)}]{Savage2006}
{Savage}, B.~D., \& {Lehner}, N. 2006, The Astrophysical Journal Supplement
  Series, 162, 134

\bibitem[{{Shapiro} \& {Field}(1976)}]{Shapiro1976}
{Shapiro}, P.~R., \& {Field}, G.~B. 1976, \apj, 205, 762

\bibitem[{{Shelton} {et~al.}(2012){Shelton}, {Kwak}, \& {Henley}}]{Shelton2012}
{Shelton}, R.~L., {Kwak}, K., \& {Henley}, D.~B. 2012, \apj, 751, 120

\bibitem[{{Shull} {et~al.}(2012){Shull}, {Smith}, \&
  {Danforth}}]{2012ApJ...759...23S}
{Shull}, J.~M., {Smith}, B.~D., \& {Danforth}, C.~W. 2012, \apj, 759, 23

\bibitem[{{Smith} {et~al.}(2011){Smith}, {Hallman}, {Shull}, \&
  {O'Shea}}]{Smith2011}
{Smith}, B.~D., {Hallman}, E.~J., {Shull}, J.~M., \& {O'Shea}, B.~W. 2011,
  \apj, 731, 6

\bibitem[{{Sommer-Larsen} {et~al.}(2003){Sommer-Larsen}, {G{\"o}tz}, \&
  {Portinari}}]{Sommer2003}
{Sommer-Larsen}, J., {G{\"o}tz}, M., \& {Portinari}, L. 2003, \apj, 596, 47

\bibitem[{{Stewart} {et~al.}(2011){Stewart}, {Kaufmann}, {Bullock}, {Barton},
  {Maller}, {Diemand}, \& {Wadsley}}]{Stewart2011}
{Stewart}, K.~R., {Kaufmann}, T., {Bullock}, J.~S., {et~al.} 2011, \apj, 735,
  L1

\bibitem[{{Stewart} {et~al.}(2017){Stewart}, {Maller}, {O{\~n}orbe}, {Bullock},
  {Joung}, {Devriendt}, {Ceverino}, {Kere{\v{s}}}, {Hopkins}, \&
  {Faucher-Gigu{\`e}re}}]{Stewart2017}
{Stewart}, K.~R., {Maller}, A.~H., {O{\~n}orbe}, J., {et~al.} 2017, \apj, 843,
  47

\bibitem[{{Sutherland} \& {Dopita}(1993)}]{Sutherland1993}
{Sutherland}, R.~S., \& {Dopita}, M.~A. 1993, The Astrophysical Journal
  Supplement Series, 88, 253

\bibitem[{{Tasker} \& {Bryan}(2006)}]{Tasker2006}
{Tasker}, E.~J., \& {Bryan}, G.~L. 2006, \apj, 641, 878

\bibitem[{{Tumlinson} {et~al.}(2017){Tumlinson}, {Peeples}, \&
  {Werk}}]{Tumlinson2017}
{Tumlinson}, J., {Peeples}, M.~S., \& {Werk}, J.~K. 2017, Annual Review of
  Astronomy and Astrophysics, 55, 389

\bibitem[{{Turk} {et~al.}(2011){Turk}, {Smith}, {Oishi}, {Skory}, {Skillman},
  {Abel}, \& {Norman}}]{Turk2011}
{Turk}, M.~J., {Smith}, B.~D., {Oishi}, J.~S., {et~al.} 2011, \apjs, 192, 9

\bibitem[{{Wakker}(2001)}]{Wakker2001}
{Wakker}, B.~P. 2001, The Astrophysical Journal Supplement Series, 136, 463

\bibitem[{{Wakker} {et~al.}(2008){Wakker}, {York}, {Wilhelm}, {Barentine},
  {Richter}, {Beers}, {Ivezi{\'c}}, \& {Howk}}]{Wakker2008}
{Wakker}, B.~P., {York}, D.~G., {Wilhelm}, R., {et~al.} 2008, \apj, 672, 298

\bibitem[{{Walch} {et~al.}(2015){Walch}, {Girichidis}, {Naab}, {Gatto},
  {Glover}, {W{\"u}nsch}, {Klessen}, {Clark}, {Peters}, {Derigs}, \&
  {Baczynski}}]{Walch2015}
{Walch}, S., {Girichidis}, P., {Naab}, T., {et~al.} 2015, \mnras, 454, 238

\bibitem[{{Welsh} {et~al.}(2007){Welsh}, {Edelstein}, {Korpela}, {Kregenow},
  {Sirk}, {Min}, {Park}, {Ryu}, {Jin}, {Yuk}, \& {Park}}]{Welsh2007}
{Welsh}, B.~Y., {Edelstein}, J., {Korpela}, E.~J., {et~al.} 2007, \aap, 472,
  509

\bibitem[{{Wood} {et~al.}(2010){Wood}, {Hill}, {Joung}, {Mac Low}, {Benjamin},
  {Haffner}, {Reynolds}, \& {Madsen}}]{Wood2010}
{Wood}, K., {Hill}, A.~S., {Joung}, M.~R., {et~al.} 2010, \apj, 721, 1397

\bibitem[{{Zheng} {et~al.}(2017){Zheng}, {Peek}, {Werk}, \&
  {Putman}}]{Zheng2017}
{Zheng}, Y., {Peek}, J.~E.~G., {Werk}, J.~K., \& {Putman}, M.~E. 2017, \apj,
  834, 179

\end{thebibliography}

\end{document}